\documentclass[10pt,journal,compsoc]{IEEEtran}
\usepackage{multirow, rotating, amsmath, wasysym}
\usepackage{tabularx}
\usepackage{wrapfig}
\usepackage{booktabs} 
\usepackage{epstopdf}
\usepackage[tight]{subfigure}
\usepackage{algpseudocode}
\usepackage{adjustbox}
\usepackage{graphicx}
\usepackage{caption}
\usepackage{algorithm}
\usepackage{soul,xcolor}
\usepackage{todonotes}

\mathchardef\Gamma="0100 \mathchardef\Delta="0101
\mathchardef\Theta="0102 \mathchardef\Lambda="0103
\mathchardef\Xi="0104 \mathchardef\Pi="0105
\mathchardef\Sigma="0106 \mathchardef\Upsilon="0107
\mathchardef\Phi="0108 \mathchardef\Psi="0109
\mathchardef\Omega="010A

\newcommand{\outline}[1]{}

\usepackage{xspace}
\usepackage{url}
\usepackage{graphicx}
\usepackage{latexsym}
\usepackage{amssymb}
\usepackage{amsfonts}
\usepackage{psfrag}
\usepackage{wrapfig}
\usepackage{comment}
\usepackage{alltt}
\usepackage{color}

\newcommand{\ie}{\emph{i.e.}\xspace}

\newcommand{\presec}{\vspace{-0.05in}}
\newcommand{\postsec}{\vspace{-0.03in}}
\newcommand{\presub}{\vspace{-0.05in}}
\newcommand{\postsub}{\vspace{-0.03in}}

\begin{document}

	\title{Monitoring Browsing Behavior of Customers in Retail Stores via RFID Imaging}

	\author{Kamran~Ali, Alex~X.~Liu,  Eugene~Chai and Karthik~Sundaresan
	\IEEEcompsocitemizethanks{\IEEEcompsocthanksitem K. Ali and A. X. Liu are with the Department Computer Science and Engineering, Michigan State University, Lansing, MI, USA, 48823.\protect\\
		E-mail: {alikamr3, alexliu}@cse.msu.edu 
		\IEEEcompsocthanksitem E. Chai and K. Sundaresan are with the NEC Labs, Princeton, NJ, USA, 08540. E-mail: {eugene, karthiks}@nec-labs.com
	}
		\thanks{Manuscript under review since October 20, 2019.}
    }
    
    \frenchspacing
    
    \IEEEtitleabstractindextext{
	
\begin{abstract}
	In this paper, we propose to use commercial off-the-shelf (COTS) \textit{monostatic} RFID devices (\ie which use a single antenna at a time for both transmitting and receiving RFID signals to and from the tags) to monitor browsing activity of customers in front of display items in places such as retail stores.
	To this end, we propose \emph{TagSee}, a multi-person imaging system based on monostatic RFID imaging.
	TagSee is based on the insight that when customers are browsing the items on a shelf, they stand between the tags deployed along the boundaries of the shelf and the reader, which changes the multi-paths that the RFID signals travel along, and both the RSS and phase values of the RFID signals that the reader receives change.
	Based on these variations observed by the reader, TagSee constructs a coarse grained image of the customers. 
	Afterwards, TagSee identifies the items that are being browsed by the customers by analyzing the constructed images.
	The key novelty of this paper is on achieving browsing behavior monitoring of multiple customers in front of display items by constructing coarse grained images via robust, analytical model-driven deep learning based, RFID imaging.
	To achieve this, we first mathematically formulate the problem of imaging humans using monostatic RFID devices and derive an approximate analytical imaging model that correlates the variations caused by human obstructions in the RFID signals.
	Based on this model, we then develop a deep learning framework to robustly image customers with high accuracy. 
	We implement TagSee scheme using a Impinj Speedway R420 reader and SMARTRAC DogBone RFID tags.
    TagSee can achieve a TPR of more than ${\sim}90\%$ and a FPR of less than ${\sim}10\%$ in multi-person scenarios using training data from just 3-4 users.
\end{abstract}
	
	\begin{IEEEkeywords}
		RF Sensing, RF Imaging, RFID, Deep Learning, Customer Activity Monitoring, Browsing Behavior, Retail Technology
	\end{IEEEkeywords}
    }

	\maketitle

	\thispagestyle{empty}
	
	\sloppy{	
		
		\presec \presec
\section{Introduction}
\postsec
%
RFIDs based activity monitoring has recently emerged as a way to monitor customer activity in physical retail stores \cite{liu2015tagbooth,han2016cbid, shangguan2015shopminer,liu2012mining, fujino2014analyzingin}.
Acquiring customer activity information is important, as the amount of time that a customer spends on browsing an item is a key indicator of the amount of interest that the customer has towards the item.
Manufacturers can use such information to improve the quality of their products, such as their visual attractiveness.
Moreover, retailers can use such information for the strategic placement of retail items \cite{karimi2012rfid, newman2003store, bloch1983shopping, bloch1989extending}.
However, existing RFID based systems for customer activity monitoring in physical retail stores \cite{liu2015tagbooth,han2016cbid, shangguan2015shopminer,liu2012mining, fujino2014analyzingin} have two key limitations.
%
First, they require physical interactions with tagged display items for detecting human interest in places such as clothing stores \cite{shangguan2015shopminer}.
%
%
Second, they do not work in multi-person environments, which are most common in reality.
In contrast, we seek to leverage COTS RFID devices for monitoring \textit{browsing activity} (\ie when there is no physical interaction between customers and the display items) of customers in retail stores. 
Such information is easy to obtain in online shopping environments by monitoring customers' online browsing behavior, such as the amount of time spent on viewing a product or the number of clicks on a product, but it is difficult to obtain in physical shopping environments.
Effective monitoring of browsing activity in physical shopping environments will not only provide useful insights on customer behavior to product manufacturers and retail store managers, but can also help to shorten the gap between online shopping and physical shopping.

In this paper, we propose \emph{TagSee}, a multi-person browsing activity monitoring system based on RFID imaging.
The hardware requirements of TagSee include a set of RFID tags and an RFID reader, both tags and the reader are COTS products.
The tags are deployed on the boundaries of the shelves.
The reader is deployed such that the customers stand in between the monitored shelves and the reader.
%
TagSee is based on the insight that when customers are browsing the items, as they stand between the tags and the reader, the multi-paths that the RFID signals travel along change, and therefore, both RSS and phase values of the RFID signals that the reader receives change as well.
Based on these variations, TagSee constructs a coarse grained image of the customers and the tags using a model-driven deep learning framework.
Afterwards, TagSee determines popularity of different item categories that are being browsed by the customers by analyzing the constructed images.
%
%
%
The key novelty of this paper is on achieving multi-person browsing behavior monitoring in front of display items by constructing coarse grained images via robust, analytical model-driven deep learning based, RFID imaging. 
%
%
TagSee works for multi-person scenarios, works for scenarios where there is no physical interaction between customers and the display items, and is device-free ($\ie$ there is no need to attach anything to shelf items or customers).
To the best of authors' knowledge, this is the first work that can monitor browsing activity of customers without cameras and without the customer touching the display items.
%
%
\begin{figure}[htbp]
	\centering
	\captionsetup{justification=centering}	\includegraphics[width=0.35\textwidth]{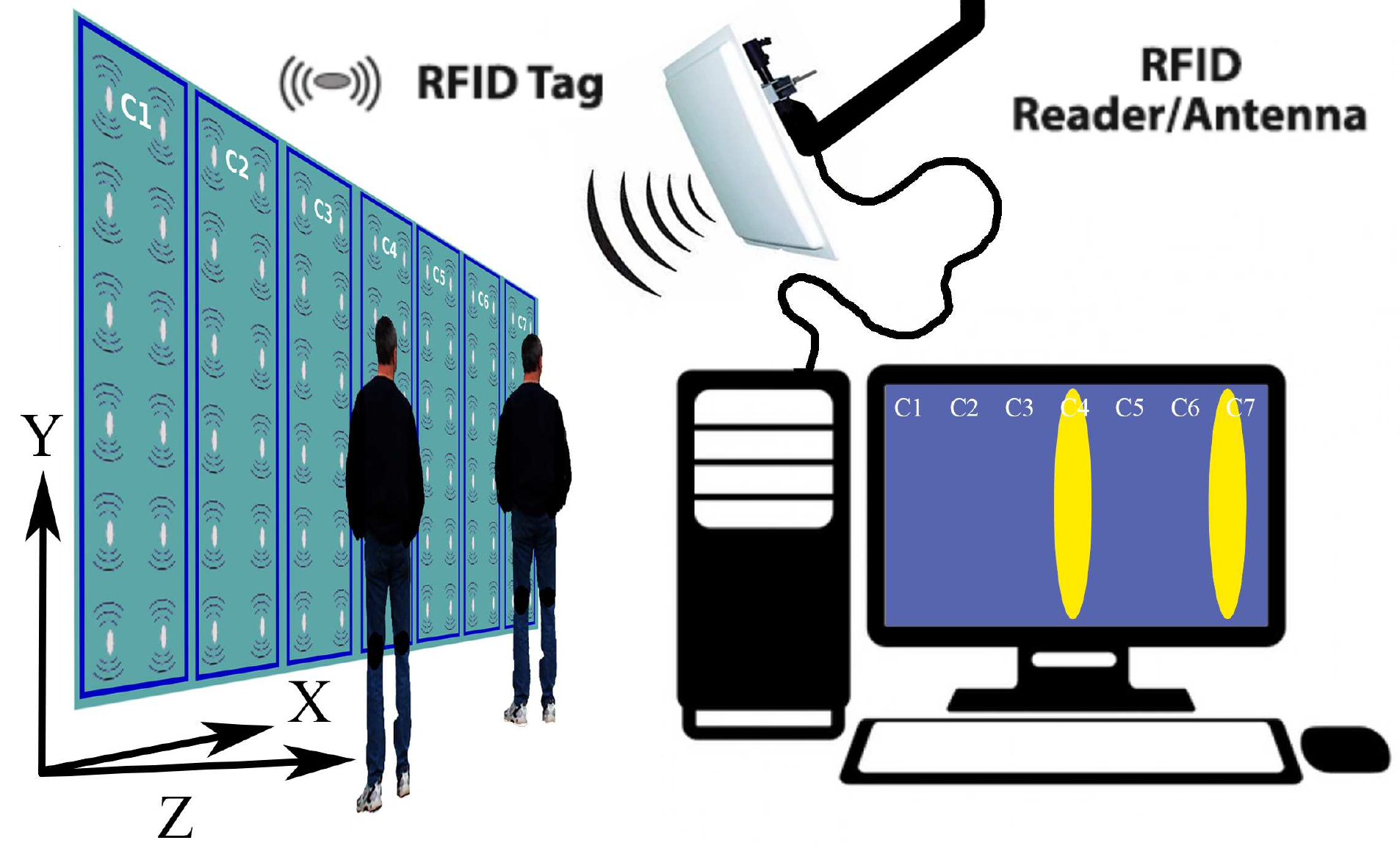}
	\vspace{-0.1in}
	\caption{Example system setup} 
	\label{fig:scenario}
	\vspace{-0.25in}
\end{figure}

%
The first technical challenge is to robustly model the relationship between the signal attenuation caused by the human obstruction in RFID signals and the images of the human obstruction.
This modeling is difficult for two major reasons.
First, the interactions between human objects and RFID signals during a browsing activity are highly complex.
%
%
%
Second, as we use \textit{monostatic} RFID readers, which use a single antenna at a time for both transmitting and receiving RFID signals to and from the tags, modeling the impact of human obstructions on RFID signals becomes even more difficult.
This is because on any reader-tag-reader (TX-Tag-RX) path, the RFID signals experience two different attenuations due to an obstruction, once when signals are sent from the reader antenna to the tag, and the second when the tag backscatters those signals towards the same reader antenna.
%
%
Employing geometrical and measurement models, such as the ones used in previous RF imaging techniques \cite{wilson2010radio,thouin2011multi,zhao2013radio,nannuru2013radio}, will entail high dependency on multiple, environment dependent and difficult to tune parameters.
Moreover, the imaging accuracy of such geometrical and measurement models based systems is not accurate and robust enough for imaging multiple customers showing interest in multiple different item categories simultaneously.
Also, as interactions between human objects and RFID signals during a browsing activity are highly complex, an accurate mathematical model is hard to derive.
%
%
%
To address these challenges, we propose a model-driven \textit{Deep Neural Networks} (DNNs) \cite{Goodfellow-et-al-2016} based RF imaging approach.
First, we mathematically formulate the problem of imaging human obstructions using monostatic RFID devices and derive an approximate analytical imaging model that correlates the variations caused by human obstructions in the RFID signals.
Second, based on the derived imaging model, we develop a DNNs based deep learning framework to robustly image customers with high accuracy. 
Our key intuition is that by training our system with the images constructed based on RFID signals when humans are browsing items, TagSee can automatically learn the underlying relationship between those images and the observed RFID signal dynamics.
%
%
%
%
%
%
%
%
%
TagSee's DNN based approach is easy to realize in practice as it is environmental and hardware independent, and the thresholds and parameters are easy to tune.
Moreover, our approach allows for robust imaging, even when customers are naturally moving to-and-fro or sideways while browsing or putting back/picking up items from a shelf.

The second challenge is to enable multi-person imaging, but without changing the training requirements.
That is, our system should only require the DNN to be trained for single-person scenarios, and it should not require the DNN to be trained for multi-person scenarios.
This is because first, it is cumbersome to train the DNN with all possible multi-person scenarios, and second, it will lead to overfitting of the DNN.
To address this challenge, we propose a spatial moving window based imaging technique to image multiple customers, who are browsing products in different columns, simultaneously.
The intuition is that, a single customer can significantly influence the RSS values of only a block of deployed tags (\ie, the ones covered by the moving window), and that multiple customers maintain a distance between themselves while browsing any shelf.
To achieve this, TagSee 
moves a window over the spatial distribution of tags, shifting it rightward, one column of tags at a time.
For each instance of the moving window, TagSee replaces the observed RSS variations for the tags lying outside the moving window with random values, which are sampled from the Gaussian distributions defined by the mean and variance of RSS variations corresponding to those tags, observed during the calibration phase (\ie, when there is no human obstruction around).
Afterwards, TagSee constructs the images corresponding to each instance of the moving window, by using the modified RSS variation vectors, which consists of changes observed in the RSS values of every tag on the shelf, as the input to the DNN.
Finally, it combines those images by applying the averaging filters to output the final image.
This DNN based approach does not require the exact locations of deployed tags and antennas to be known in advance, which makes it simple to deploy in practice.

The third challenge in TagSee is to robustly quantify the variations introduced by human obstructions in the RSS values of the deployed tags.
This is necessary because TagSee uses these RSS variations as inputs to its DNN for imaging obstructions.
Anomalous variations in RSS values of different tags occur frequently during browsing behavior, either due to fading loss from constructive/destructive interference of RFID signals due to multipath effects, or due to the measurement noise of the RFID reader.
To address this challenge, we leverage the \textit{frequency hopping} (FH) capability of the multi-frequency UHF RFID hardware, which operates in the 902-928 MHz frequency range (divided into 50 closely spaced subcarriers).
%
%
First, in scenarios where some part of RFID spectrum is under fade while a reader attempts to interrogate a tag, FH capability allows the reader to interogate that tag on stronger subcarriers, which helps TagSee gather enough measurements per tag per second, required for robust and accurate image reconstruction.
Second, since the subcarriers are closely spaced in the frequency range 902-928 MHz, the impact of disturbances created by a human subject on the RSS values corresponding to a certain tag, is similar across most subcarriers because transmitted power in all subcarriers are the same.
Based on this intuition, TagSee robustly estimates the variations in RSS values of different deployed tags by taking the median of the RSS variations observed over multiple subcarriers.
This reduces anomalies in the variations of RSS values, leading to significantly reduced distortions in the constructed images. 

%
%
We implement TagSee system using a Impinj Speedway R420 reader and SMARTRAC DogBone RFID tags.
%
We attach RFID tags in a distributed and orderly fashion, just like a mesh, along the boundaries of a shelves, while covering all column-wise item categories.
We call such tags which are attached to the shelves as \textit{Static Tags}.
In any monitoring scenario consisting of $A$ number of RFID antennas, there are $A\ast K$ unique TX-Tag-RX links for $K$ tags deployed along the boundaries of the shelf (so number of links $M = K$ in our case).
We use Impinj Speedway R420 RFID readers that are capable of reading upto $\sim 450-500~tags/s$.
This allows us to gather enough RSS and phase information from the deployed tags to achieve smoother monitoring of browsing behavior.
We create real-life scenarios to perform comprehensive experiments involving 10 different human subjects with IRB approval, and then evaluate the performance of TagSee on this dataset.
%
%
Our experimental results show that, on average, TagSee can achieve a TPR of more than ${\sim}90\%$ and a FPR of less than ${\sim}10\%$ in multi-person scenarios using training data from just 3-4 users.
		\presec\presub
\section{Related Work} \label{sec:relatedwork}
\postsec
%
%

%
\subsection{Radio Tomographic Imaging (RTI)}
\postsub
Previously proposed RTI approaches, which are closest to our work, are either based on sensor networks \cite{wilson2010radio,thouin2011multi} or bistatic passive RFID (pRFID) systems \cite{wagner2012passive}.
In the sensor networks based approaches \cite{wilson2010radio,thouin2011multi}, each node deployed around a monitored area is capable of both transmitting and receiving RF signals, independently, and for any TX-RX pair, there is exactly one communication link which gets affected by an obstruction.
%
%
In contrast, the bistatic passive RFIDs based system proposed in \cite{wagner2012passive} entails two communication links per tag read, i.e. a TX-Tag link from the TX antenna which interrogates the tag, and a Tag-RX link where the RX antenna receives the response back from the tag (we refer to RFID links as TX-Tag-RX in rest of the paper).
%
%
However, both of the aforementioned RTI scenarios are similar, because in each case, RF signals experience attenuation due to an obstruction only once on their way to the receiver side.
Moreover, all previous RTI schemes have an inherent issue of high dependence on multiple, difficult to tune parameters, such as the parameters corresponding to different types of geometrical and measurement models, which are used to capture the effects of attenuation due to human obstructions \cite{wilson2010radio,thouin2011multi}.
First, these parameters are often highly dependent on experimental scenarios and the hardware being used.
Second, these parameters have to be tuned manually, which is time consuming, and often requires intensive calibration for each different deployment scenario.
Third, inefficient tuning of these parameters leads to unstable and ineffective imaging results.
Moreover, all the previous RF imaging approaches require the exact locations of all RF nodes to be known in advance.
%
%
These limitations make previously proposed RTI techniques difficult to realize.
Compared to these previous RTI works, we develop a model-driven DNNs scheme for monostatic RFID systems, which gets rid of manual setting of difficult to tune RFID channel parameters and enables accurate multi-person imaging using monostatic passive RFID hardware.

\presec
\presec\presec
\subsection{Customer Behavior Monitoring using RFIDs}
\postsub
RFID based techniques for human activity tracking in physical retail \cite{liu2015tagbooth,han2016cbid, shangguan2015shopminer,liu2012mining, fujino2014analyzingin} utilize variations in received signal strength (RSS) and phase values of RFID signals to monitor customer behavior.
However, most RFID based behavior monitoring techniques such as \cite{liu2015tagbooth,han2016cbid, shangguan2015shopminer} only focus on clothing stores. Moreover, the current RFID based techniques are highly parameter dependent and fail to work well in multi-person scenarios.
%
%
Moreover, these techniques require all retail items in a shelf to be tagged with RFID tags, a requirement which is often not satisfied in many retail stores.
To the best of our knowledge, there is no prior work that can monitor customer browsing activity without using cameras or the requirement of physically touching retail items tagged with RFIDs.

\presec\presec
\presec
\subsection{Customer Behavior Monitoring using Cameras} 
\postsub
Several camera based solutions to customer behavior analytics exist in literature \cite{connell2013retail, kobres2016whole, burke2017system, liciotti2014shopper, liu2015customer, dconsumer, popa2010analysis, popa2011kinect, pereira2013motion}.
	A key advantage of RFID based solutions compared to dense deployment of cameras is that RFID based solutions are not privacy intrusive. 
	In the past, multiple privacy concerns related to camera based solutions have been brought to light, for example with the `Amazon Go' stores that use dense deployment of cameras to monitor customers \cite{amazon1, amazon2}.
	However, please note that even though our system has a privacy advantage, we do not seek to replace any existing camera based solutions.
	Instead, we envision that TagSee can be integrated with existing camera based techniques for improved customer behavior analytics and enhanced shopping experience solutions such as automatic check-out \cite{enhanced-impinj}.
	Such integration of RFID and camera based solutions will also help reduce privacy concerns by allowing the retail stores to enable smart solutions with less number of cameras. 
	The goal of this paper is to advance the state-of-the-art in the emerging field of RFIDs based customer behavior analytics in retail stores by leveraging off-the-shelf RFID readers and tags for monitoring browsing activity of customers (i.e. when there is no physical interaction between customers and the display items).
	To the best of our knowledge, there is no prior work that can monitor customer browsing activity without using a camera or without the requirement of first attaching tags on the retail items with RFIDs and then physically touching those items.

		\presec \presec\section{System Overview} \label{sec:overviewseewsee} 
\postsec

%
%

\subsection{Monostatic Passive RFIDs} \postsub \label{sec:overview}

In a monostatic pRFID system, a \textit{reader} transmits continuous wave signals to interrogate tags deployed in its proximity, and then receives backscattered signals from those tags which contain their unique IDs.
In this work, use industrial standard, EPC Global Class 1 Generation 2 (C1G2) RFID \cite{epcglobal2008radio} compatible, Ultra-High-Frequency (UHF) pRFID tags and Impinj Speedway R420 monostatic RFID readers to realize TagSee.
The RFID readers we use operate in the frequency range 902-928 MHz, through a \textit{frequency hopping} (FH) mechanism, where the frequency range is divided into 50 subcarriers (\ie 902.75 - 927.25 MHz with an interval of 0.25 MHz), which are randomly hopped between during each interrogation cycle. 
This FH capability reduces interference between nearby RFID readers, and leads to robust and reliable interrogation of tags in cases where some part of the spectrum is under fade.
Also, these readers are equipped with multiple antennas.
The query-response communication corresponding to each antenna is multiplexed in time, where each antenna interrogates the tags in an alternating manner.
In each query-response communication, the EPC C1G2 compatible RFID tags respond to RFID signals from a reader, through a random access collision avoidance technique called \textit{slotted} ALOHA \cite{kawakita2006anti}.
The RFID readers we use are capable of reading upto $\sim500~tags/s$, which allows for gathering enough RSS and phase information from the deployed tags, required for smoother and near real-time monitoring of customers' browsing behavior.

\begin{figure*} [htbp]
	\vspace{+0.1in}
	\centering
	\captionsetup{justification=centering}
	\includegraphics[width=0.9\textwidth]{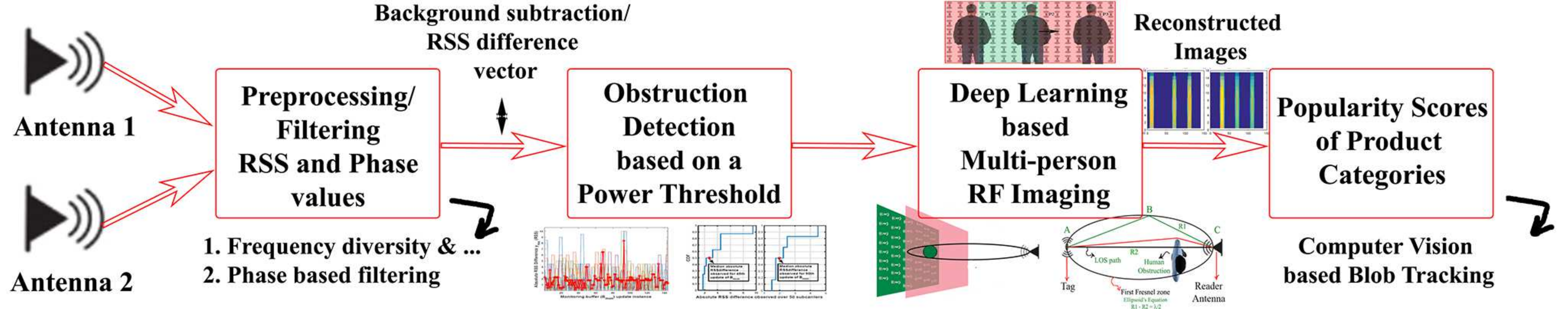}
	\caption{High level flow diagram of TagSee's monitoring mode } 
	\label{fig:system}
	\vspace{-0.2in}
\end{figure*}

\presub\presub\presub
\subsection{TagSee's Imaging Infrastructure}
\postsub
TagSee requires items to be placed in column-wise categories on shelves.
TagSee also assumes that RFID tags are already attached in a distributed and orderly fashion, just like a mesh, along the boundaries of different shelves, while covering all item categories.
We call such tags which are attached to the shelves as \textit{Static Tags}.
%
%
Although, TagSee does not require the items placed in the shelves to be tagged, yet, for the sake of completeness, we call such tags as \textit{Dynamic Tags}, because their positions can change with time or they can even disappear from their shelves in cases where those items get purchased.
%
%
%
%
%
%
In any monitoring scenario consisting of $A$ number of RFID antennas, there are $A\ast K$ unique TX-Tag-RX links for $K$ tags deployed on a shelf (so number of links $M = K$ in our case).
To monitor activity of customers in front of a shelf or a display item, an RFID \textit{reader} transmits continuous wave signals to interrogate tags deployed in its proximity, and then receives backscattered signals from those tags which contain their unique IDs.
Afterwards, the RSS and phase values corresponding to these tags are leveraged to image any customers standing between the tags and the reader.
%
%
%
%
%
Our analytical model based monostatic RFID imaging approach requires the locations of all static tags and RFID reader antennas to be known in advance. 
However, our final deep learning based imaging approach does not have any such requirement.
%

\presub\presub
\subsection{TagSee's System Level Overview}
\postsub

TagSee consists of two working modes, namely \textit{calibration} mode and \textit{monitoring} mode. 
In calibration mode, TagSee reads RSS and phase values from the deployed tags when there is no human obstruction in the monitored area.
These calibrated RSS and phase values are used for background subtraction during the \textit{monitoring} mode.
In \textit{monitoring} mode, TagSee constructs images after processing the RSS and phase values it reads from the deployed tags.
Figure \ref{fig:system} shows a system level diagram of TagSee for \textit{monitoring} mode.
In the first step, the raw RSS and phase values obtained from the deployed tags are fed into a pre-processing module.
The pre-processing module first applies a combination of moving average and moving median filters on streaming RSS and phase data, and then subtracts the calibrated RSS and phase values from the incoming filtered RSS and phase values, respectively.
Afterwards, it filters out the anomalous variations in RSS values by applying \textit{phase difference} and \textit{frequency diversity} based filtering techniques.
In the second step, the power of resultant RSS difference vectors obtained from pre-processing module is checked with a threshold, to determine the existence of obstruction in front of the deployed tags.
If any obstructions are detected, the RSS difference vectors are then fed into a DNN based multi-person RFID imaging module, which then constructs coarse-grained images of the detected human obstructions.
Finally, TagSee applies a computer vision based blob tracking technique on the constructed images, to monitor the browsing activity near different items, by determining the popularity scores of different item categories.
Next, we discuss how TagSee processes the RSS and phase values during its two working modes.

\presec\presec
\section{Preprocessing RSS and Phase} \label{sec:preprocess}
\postsec

\smallskip\noindent\textbf{Received Signal Strength (RSS):} Monostatic RFID channel is a double fading channel, \ie each fade is experienced twice, once in the forward link and once in the reverse link.
A typical RFID use case scenario involves indoor multipath environment, where each link consists of a line-of-sight (LOS) path and a few major reflections.
For Mono-static RFID readers, for a given subcarrier of wavelength $\lambda$, the received power $P_{r}^{\{k,a\}}$ at the reader antenna $a$, of the backscattered signal from tag $k$ located at distance $d$ from the reader, can be approximated in terms of transmit power $P_{t}$ for a free space scenario as:
\vspace{-0.05in}
\begin{equation}
P_{r}^{\{k,a\}} = P_{t} G_a^2 G_{k}^2 T_{b}^{\{k\}} \cdot (\frac{\lambda}{4\pi d})^4
\label{eqn:solution4}
\vspace{-0.05in}
\end{equation}
where $G_{a}$ and $G_k$ are $a^{th}$ reader antenna and $k^{th}$ tag antenna gains, respectively \cite{nikitin2008antennas}. $T_{b}^{k}$ is the backscatter or modulation loss of the $k^{th}$ tag. Next, we assume that $P_{t}$, $G_k$, $G_a$ and $T_{b}^{\{k\}}$ remain constant for a tag-antenna pair, and then re-write a simplified logarithmic relation for RSS at $k^{th}$ tag as follows:
\vspace{-0.04in}
\begin{equation}
RSS^{\{k\}}~(dBm) = A_{0}^{\{k\}} + 10 \cdot \beta^{\{k\}}\log[(\frac{\lambda}{4\pi d})]
\label{eqn:solution6}
\vspace{-0.03in}
\end{equation}
where $A_{0}^{\{k\}}$ is assumed to be a constant for a specific environment, and the value of $\beta^{\{k\}} = 4$ in case of LOS free space path loss scenario depicted in Eq. (\ref{eqn:solution4}).
In reality, $\beta^{\{k\}}$ is dependent upon indoor multipath environment and shadowing effects due to obstructions.
We use Eq. (\ref{eqn:solution6}) while formulating our analytical RFID imaging approach.

\smallskip\noindent\textbf{Phase:} For RFID propagation environment involving monostatic readers, the phase information of a received signal, received from $k^{th}$ tag, provided by the reader can be written as $\phi = mod(\phi_p + \phi_o + \phi_{b,k}, 2\pi)$, 
where $\phi_p = 2\kappa d + \phi_m$ ($\kappa = \frac{2\pi f}{c}$, $f$ = signal frequency and $\phi_m$ phase contribution due to constructive/destructive interference due to multipaths), $\phi_o$ is the phase offset which includes phases of the cables and other reader and antenna components, and $\phi_{b}^{\{k\}}$ is the backscatter phase of the $k^{th}$ tag modulation.
Next, we discuss the techniques we use in TagSee's \textit{calibration} and \textit{monitoring} working modes to pre-process RSS and phase data before feeding it to the imaging module.

\presub \presub\presub
\subsection{Calibration Mode}
\postsub
In calibration mode, TagSee reads RSS and phase values from the deployed tags when there is no human obstruction in the monitored area.
These calibrated RSS and phase values are used for background subtraction during the \textit{monitoring} mode.
During calibration, TagSee keeps reading the RSS and phase values for $t_{cal} \approx 2~minutes$, to ensure that it receives enough readings from each of the deployed tags and carrier frequencies used by the reader.
For $F$ frequencies and $K$ tags, TagSee records $FK$ RSS and phase vectors, per RFID antenna.
In the end, TagSee applies a combination of moving average and moving median filters (window size = 5) to all $FK$ number of RSS and phase vectors, calculates the median of each of those $FK$ filtered vectors, and records $F \times K$ dimensional RSS and phase calibration matrices ($M_{rss}^{cal}$ and $M_{phase}^{cal}$), corresponding to each of the $A$ antennas, for background subtraction during image construction in the monitoring mode.

\presub\presub
\subsection{Monitoring Mode}
\postsub
In \textit{monitoring} mode, TagSee constructs images after processing the RSS and phase values it reads from the deployed tags.
During this mode, TagSee keeps reading the RSS and phase values, while maintaining the streaming values in a buffer $B_{mon}$ for batch processing.
In our experiments, we observed that every user takes at least $3-4~secs$ while browsing a certain item category, which amounts to approximately $N_{mon} = 2000$ RSS and phase readings. 
Therefore, during monitoring mode, TagSee maintains latest $N_{mon} = 2000$ readings in buffer $B_{mon}$.
$B_{mon}$ is updated with new readings every $t_{mon} = 1~secs$, which amounts to approximately 450-500 readings.
As all tags contend for the medium through a random access protocol, TagSee might not receive readings for some tag-frequency pairs in a period of $3-4~secs$.
For utilizing frequency hopping capability of RFID readers efficiently during the frequency diversity based filtering process that we discuss later on, TagSee waits until it gets enough readings from the deployed tags for multiple different frequencies.
We experimentally observe that the above values of $N_{mon}$ and $t_{mon}$ allow for recording enough readings for robustly constructing reasonable images.
%
%
However, we also observed that during a browsing activity, it often happens that the data obtained during a certain time window does not contain data from all the deployed tags.
Advanced \textit{matrix completion} algorithms \cite{recht2011simpler, keshavan2010matrix} can be used to interpolate missing RSS data, however it will significantly increase the computational complexity.
Therefore, in the case where TagSee does not find any reading in $B_{mon}$ for a certain tag-frequency pair, it replicates the calibrated RSS and phase readings corresponding to that tag-frequency pair.
Finally, TagSee applies a combination of moving average and moving median filters (window size = 5) to all $FK$ number of RSS and phase vectors contained in $B_{mon}$ corresponding to each antenna.
Next, we describe how TagSee leverages the frequency diversity and phase difference to calculate a robust estimate of RSS difference vector $\mathbf{y_{rss}}$ for each RFID antenna.

\smallskip\noindent\textbf{Phase Difference based Filtering:} To filter RSS based on phase difference, TagSee first calculates phase difference vector for each of the $FK$ phase vectors contained in $B_{mon}$ by subtracting the calibrated phase values in $M_{phase}^{cal}$ corresponding to each of the $FK$ frequency-tag pair, to obtain $B'_{mon}$. 
Next, TagSee leverages the concept of Fresnel zones \cite{hristov2000fresnal} to filter out RSS values in $B_{mon}$.
The first Fresnel zone determines the LOS path between two RF nodes, and encompasses most of the RF wavefronts which contribute significantly to RF propagation.
Therefore, if the first Fresnel zone is clear of obstructions, we can assume LOS communication between a tag and its reader.
For monostatic RFIDs, the phase difference between the direct LOS path, and any other RF propagation path lying within the first Fresnel zone, can be at max $2\pi$, as show in \ref{fig:fresnel_zones_1b}, which corresponds to one wavelength.
Assuming that the phase values obtained for each tag-frequency pair during calibration phase ($M_{phase}^{cal}$) correspond to the direct LOS paths between those tags and their corresponding antenna, TagSee discards all RSS values in $B'_{mon}$, for which the absolute difference between their phase readings and the corresponding calibrated values in $M_{phase}^{cal}$, is less than $\phi_{mon} = \pi/4$.

\begin{figure*} [htbp]
	\centering
	\captionsetup{justification=centering}
	\subfigure[ \noindent{$y_{rss}$ filtered using phase values \& frequency diversity (bold-red) is plotted over RSS differences obtained for 50 subcarriers}]{
		\label{fig:filtering1a}
		\includegraphics[width=0.55\textwidth]{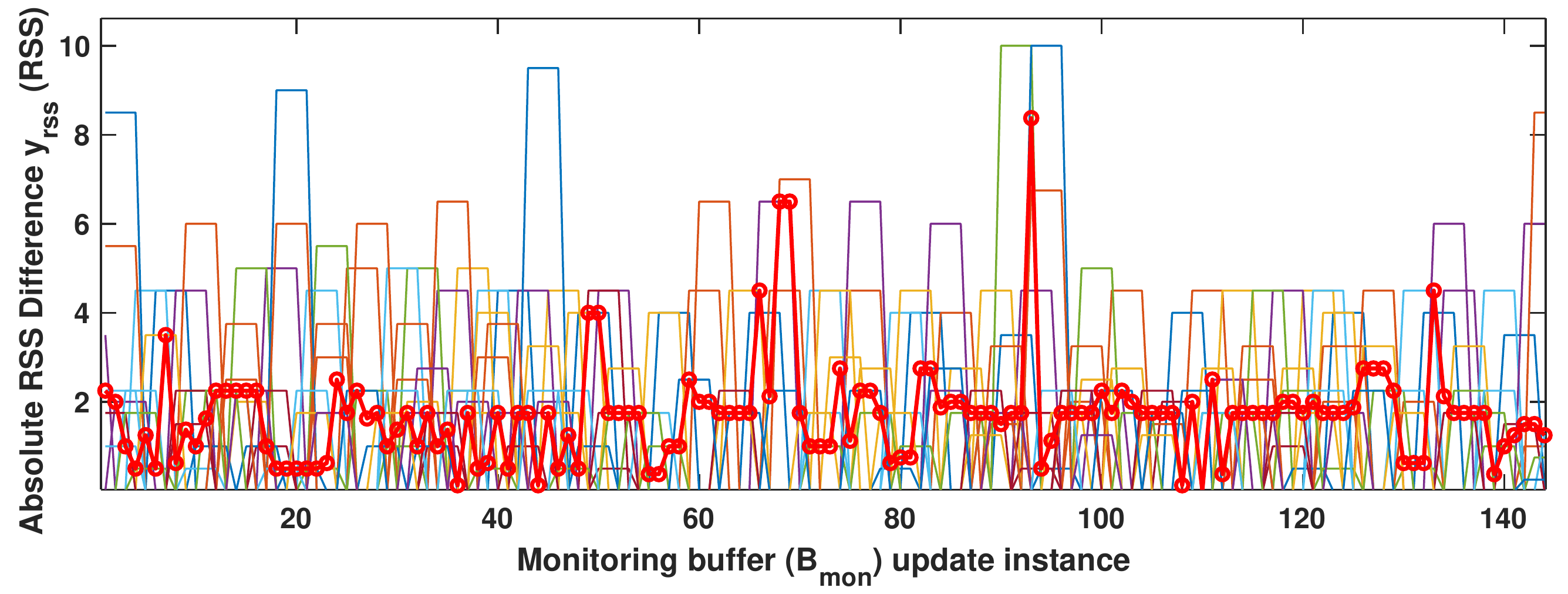}%
	}
	\hspace{0.2in}
	\subfigure[ \noindent{CDF of absolute RSS difference values obtained from 46th tag for 50 subcarriers}]{
		\label{fig:filtering1b}
		\includegraphics[width=0.297\textwidth]{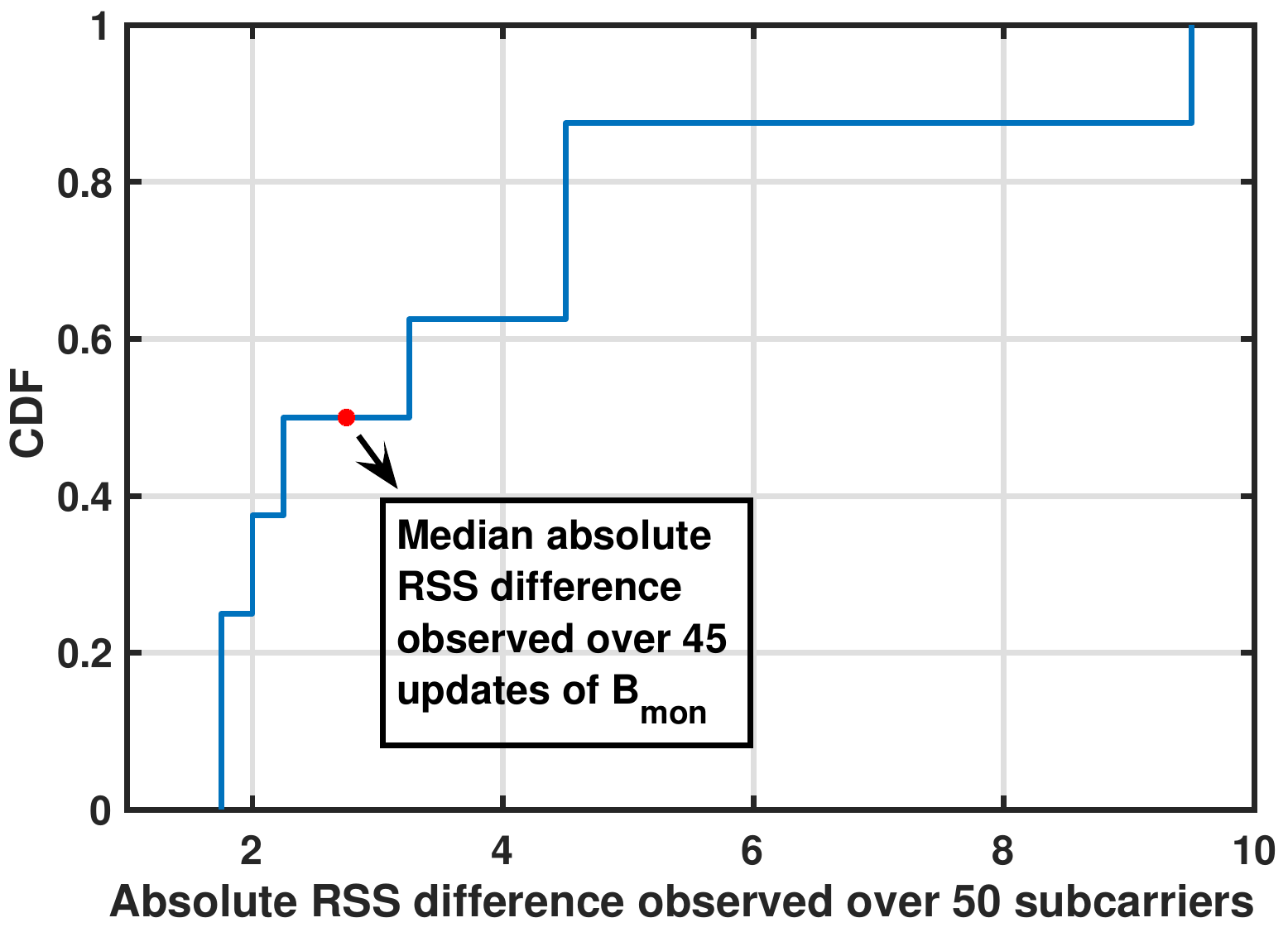}%
	}
		\vspace{-0.1in}
	\caption{Phase and frequency diversity based filtering for a tag obstructed by a human (head \& arms allowed to move)} 
	\label{fig:filtering1}
	\vspace{-0.2in}
\end{figure*}
\begin{figure} [htbp]
	\vspace{-0.13in}
	\centering
	\captionsetup{justification=centering}
	
	\subfigure[ \noindent{Image plane intersecting a Fresnel zone}]{
		\label{fig:fresnel_zones_1a}
		\includegraphics[width=0.34\textwidth]{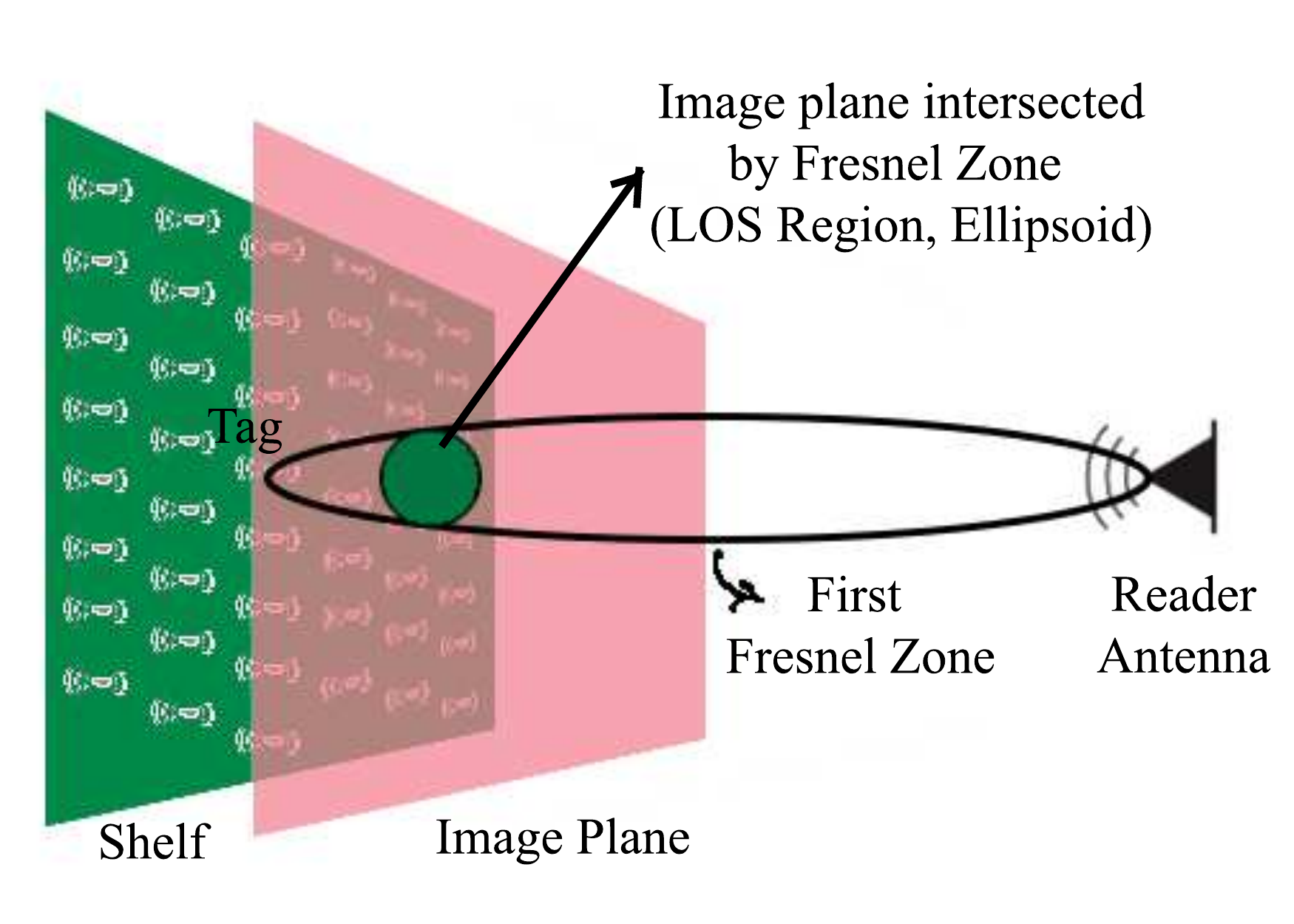}%
	}
	\subfigure[ \noindent{Impact of human obstruction on paths}]{
		\label{fig:fresnel_zones_1b}
		\includegraphics[width=0.34\textwidth]{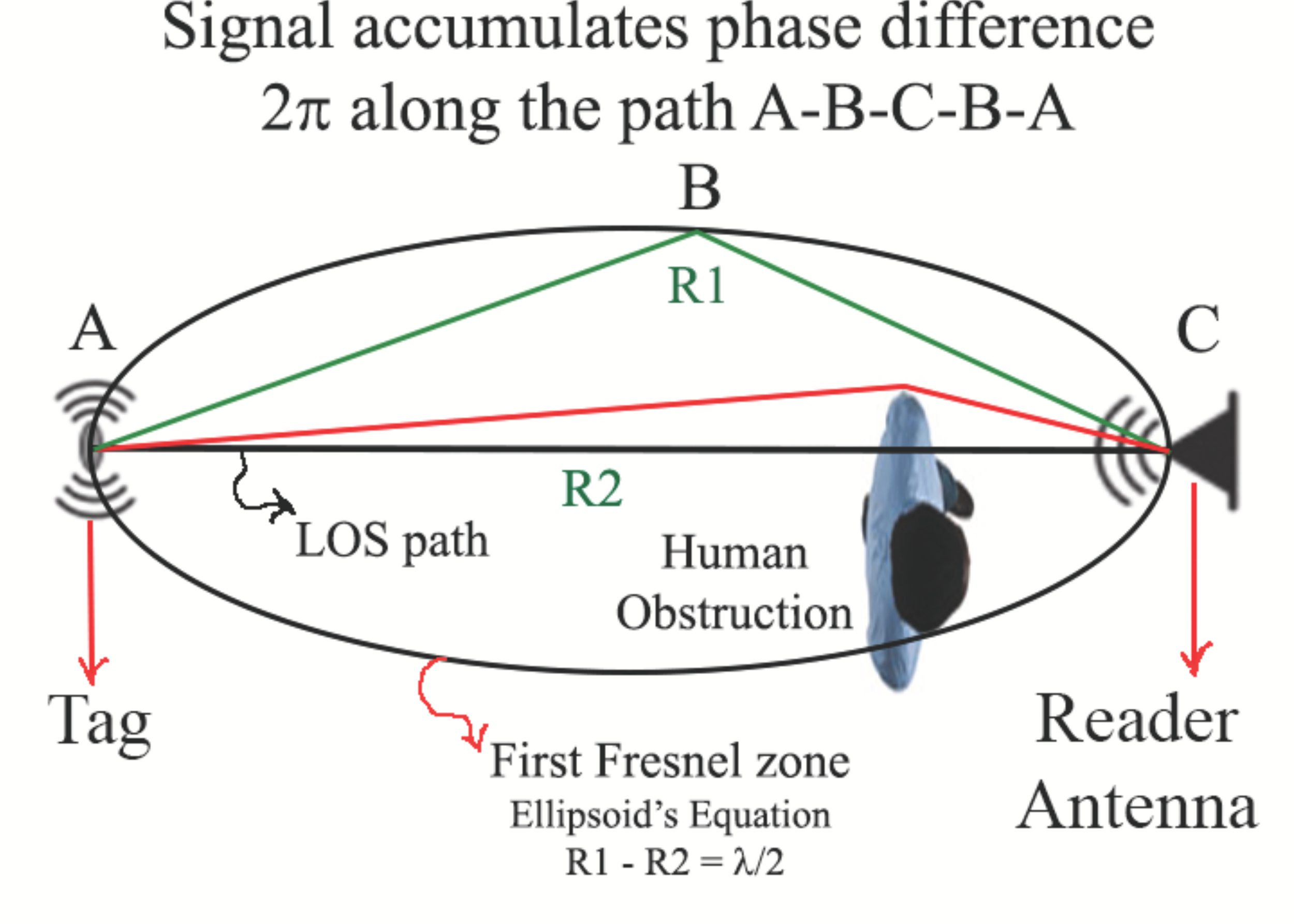}%
	}
	\vspace{-0.08in}
	\caption{Intuitions behind using the concept of First Fresnel zones \cite{hristov2000fresnal} for phase based filtering \& imaging} 
	\label{fig:fresnel_zones_1}
	\vspace{-0.24in}
\end{figure}

We chose $\phi_{mon} = \pi/4 = 0.125\times 2\pi$ because we want to select those RSS values for imaging, which correspond to the scenarios where a human subject obstructs at least 6.25\% of the Fresnel zone.
As R420 reader only provides phase values in the range $0-2\pi$, phase wraps may occur, which can lead to improper phase difference based filtering.
However, first of all, the phase difference between two RF propagation paths lying within the first Fresnel zone can be at max $2\pi$.
Second, neither the tags nor the RFID reader antennas move during the experiments.
%
%
Therefore, TagSee assumes that phase difference between the phase values read during calibration and the ones read during monitoring mode remains below $2\pi$, and does not take into account phase wraps of over $2\pi$.
Note that, a few longer paths, \ie paths outside the first Fresnel zone, may exist at any point in time.
However, we assume that the phase values corresponding to those paths are filtered out by aforementioned moving average and moving median filtering.
After applying this filter on readings obtained from each antenna separately, TagSee calculates the median of each of those $FK$ filtered vectors, and records $F \times K$ dimensional RSS matrix ($M_{rss}^{mon}$) corresponding to each of the $A$ antennas.

\smallskip\noindent\textbf{Frequency Diversity based Filtering.}
To calculate $\mathbf{y_{rss}}$, TagSee first calculates the absolute difference between calibrated and monitored RSS matrices, \ie $Y_{rss} = |M_{rss}^{cal}$ - $M_{rss}^{mon}|$, which gives an $F \times K$ dimensional RSS difference matrix.
Next, TagSee leverages the frequency diversity of the RFID system we use, to reduce anomalous variations in $Y_{rss}$.
Since the frequencies which RFID reader randomly hops between are closely spaced in the frequency range 902-928 MHz, the RSS difference observed for a tag over these multiple closely spaced frequencies, due to impact of human obstruction in a small time window of $3-4~secs$, is similar, given that the trasnmit power in each subcarrier is the same).
%
%
Therefore, TagSee takes the median of RSS difference values in $Y_{rss}$ over $F=50$ different frequencies, to obtain the $K$ dimensional RSS difference vector $\mathbf{y_{rss}}$, which is then fed into the imaging module, after every $t_{mon} = 1~secs$.
Again, this filtering is performed for each antenna.
Figure \ref{fig:filtering1} shows the filtered RSS difference values, plotted in bold-red color over the RSS difference values corresponding to all 50 subcarriers, for a slightly blocked tag by a human standing at the same place, but allowed to move head and arms.
We can observe that the filtered RSS difference values remain at a relatively stable level over time, as compared to the unfiltered RSS difference values showing the varying nature of individual subcarriers.

\smallskip\noindent\textbf{Power based filtering before RFID Imaging.} Before feeding the filtered $\mathbf{y_{rss}}$ into imaging module, TagSee checks if the changes observed in RSS values of different tags deployed along a shelf are significant enough to reveal existence of an obstruction. 
If there are no significant changes observed, then imaging should be skipped, which reduces unnecessary computations during image construction.
To achieve this, TagSee compares the power contained in the $\mathbf{y_{rss}}$ vector with a simple to tune threshold $(\mathbf{P_{rss}} =10)$, and discards any  $\mathbf{y_{rss}}$ vectors which do not meet this threshold.
%
%
Note that, the aforementioned phase based filtering of RSS values, along with this power threshold based filtering of $\mathbf{y_{rss}}$ vectors, filter out the variations in RFID channel which do not correspond to a obstruction, which leads to more accurate imaging.
%

\presec\presec
\section{Analytical RFID Imaging Approach} \label{sec:baseline}
\postsec
In this subsection, we propose an analytical imaging approach for monostatic RFID systems, which forms the basis of our proposed \textit{deep neural networks} (DNNs) based imaging technique in \S \ref{sec:neuralnet}.
%
%
We explain our model using Fig. \ref{fig:fresnel_zones_1a}.
The green surface represents the surface of the shelf on which RFID tags are deployed.
In a typical browsing scenario, customers will be standing closer to the shelf.
Our aim is to image the customers while they stand close to the shelf for browsing the items displayed on that shelf.
To develop our analytical model, we first erect an imaginary image plane at a point $P_z$ along Z-axis, parallel to the X-Y-axes as shown in Fig. \ref{fig:fresnel_zones_1a}.
Next, we divide the image plane into \textit{voxels}, such that there are $p_x$ \textit{voxels} between each pair of tags along X-axis, $p_y$ \textit{voxels} between each pair of tags along Y-axis, and the coordinates of bottom leftmost \textit{voxels} and top rightmost \textit{voxels} are $(0,~0)$ and $(Max_x,~Max_y)$, respectively ($p_x = p_y = 5$ in our current implementation, with inter-tag distance of $5~inches$ along both axes).
%
%
%
Our imaging problem can then be represented as a linear system of equations, where the changes in RSS of different links $\mathbf{y_{rss}}$ can be written in terms of the changing attenuations $\mathbf{x}$ and an $M \times N$ weight matrix $\mathbf{W}$ (where $M = $ Number of links, $N = $ Number of \textit{voxels} in the image plane), specifying the contributions of each link towards the changes observed in attenuation at each different \textit{voxel}, as 
 $\mathbf{y_{rss}} = \mathbf{W} \mathbf{x} + \mathbf{n}$,
where $\mathbf{n}$ corresponds to fading and measurement noise.
The goal is to solve the system for each antenna separately, using the RSS difference vectors for each tag-antenna pair.

For the system represented by  $\mathbf{y_{rss}} = \mathbf{W} \mathbf{x} + \mathbf{n}$, we first model the weight matrix $\mathbf{W}$ by employing the concept of \textit{Fresnel Zones} \cite{hristov2000fresnal} between two nodes of a RF link, and use imaginary ellipsoids centered at the locations of different RF nodes in the network, to determine weights of different \textit{voxels} in the image plane covering the monitored area.
The intuition is, that these ellipsoids determine the LOS path (typically chosen to be the first Fresnel zone) of each respective link, and if a \textit{voxel} is intersected by the LOS path of a link, it will be assigned more weight as compared to the \textit{voxel} which does not fall in LOS.
Figure \ref{fig:fresnel_zones_1a} shows an example scenario.
To construct image from the RSS difference values obtained for all the links (\ie $\mathbf{y_{rss}}$), the most straightforward way is to calculate the least-squares solution as $\mathbf{x_{LS}} = \mathbf{P}\times\mathbf{y_{rss}}$, where $\mathbf{P} = (\mathbf{W}^T\mathbf{W})^{-1}\mathbf{W}^T$. 
However, the matrix $\mathbf{W}$ is not full rank in case of imaging systems, which makes imaging an \textit{ill-posed inverse problem}. 
To handle ill-posedness, we use \textit{Tikhonov Regularization} approach (such as the one proposed in \cite{wilson2010radio} for minimizing objective function of the original problem. 
%
%
%
%
%
%

\smallskip\noindent\textbf{Estimating Change in Path Loss:} 
Monostatic RFID communications make the task of estimating weight matrix $\mathbf{W}$ more difficult, because signals experience attenuations multiple times before reaching back to the same receiver, \ie once when signals are sent to the deployed tags, and second when those tags backscatter those signals back to the reader antenna.
%
%
Assuming that both forward and reverse channels of a TX-Tag-RX link are symmetric, we assume an imaginary set of symmetric ellipsoids (approximate LOS regions) between each tag and reader antenna pair. 
Intuitively, the image plane cuts the Fresnel zones between each TX-Tag-RX link.
Weights are assigned to each \textit{voxel} of an image plane, based on whether they fall inside the imaginary ellipsoid of the TX-Tag-RX link or not.
%
%
%
Given the value of free space path loss exponent for a TX-Tag-RX link corresponding to each tag $k$ is 4, we can assume that during calibration phase, the initial value of $\beta_{init}^{\{k\}} = 4 + \phi_{init}^{\{k\}}$. 
Similarly, let us assume that for each new update of the buffer $B_{mon}$ during the monitoring mode, the new $\beta_{new}^{\{k\}} = 4 + \phi_{new}^{\{k\}}$. 
Both $\beta_{init}^{\{k\}}$ and $\beta_{new}^{\{k\}}$ are unknown here, however, as will show next, only the difference $\beta_{new}^{\{k\}} - \beta_{init}^{\{k\}}$ is required. 
At any time instance, we can write this change in path loss exponent in terms of the change in RSS for tag $k$ (i.e. $y_{rss,k}$) at the RFID reader antenna as follows: 
\begin{equation}
\bigtriangleup\beta^{\{k\}} = \beta_{new}^{\{k\}} - \beta_{init}^{\{k\}} = \frac{y_{rss,k}~(dBm)}{ 10\cdot\log[(\frac{\lambda_{avg}}{4\pi d})]}
\label{eqn:solution7}
\vspace{-0.05in}
\end{equation}

\begin{figure*} [htbp]
	\centering
	\captionsetup{justification=centering}
	\includegraphics[width=1\textwidth]{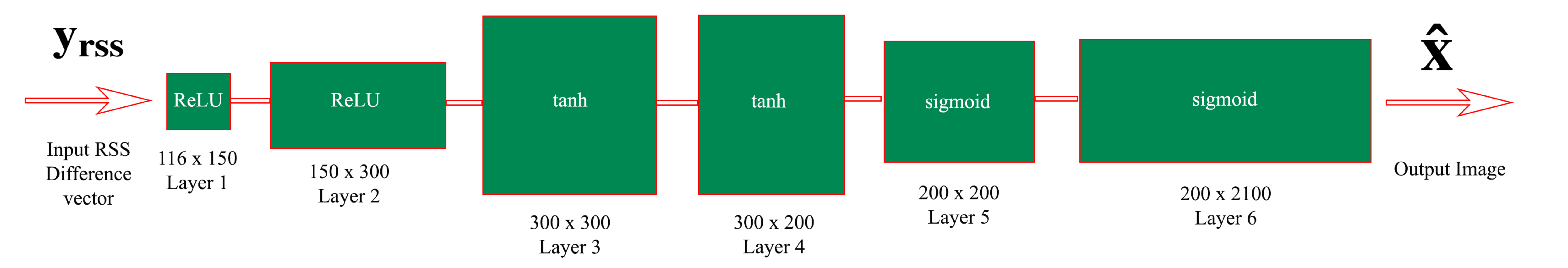}
		\vspace{-0.05in}
	\caption{DNN architecture for $K = 116~tags$ , $k_x = 29~tags$, $k_y = 4~tags$ and inter-tag distance of $5~inches$ along both axes} 
	\label{fig:neuralnetwork}
	\vspace{-0.22in}
\end{figure*}

As the RFID reader uses multiple frequencies, we use average of those frequencies and choose $\lambda = \lambda_{avg}$ for simplicity. Given the fact that these frequencies are close to each other, the choice of this value of $\lambda$ does not hurt the results of this scheme. 
After calculating $\bigtriangleup\beta$, TagSee assigns weights to each \textit{voxel} on the image plane as: 
\[
w_{kj}= 
\begin{cases}
\frac{1}{d^{(4-\bigtriangleup\beta})},& \text{if } d_{kj,1} + d_{kj,2} < d + \Theta\\
0,              & \text{otherwise}
\end{cases}
\]
Here $d$ is distance between the reader and tag for a link $k$, $d_{kj,1}$ is the distance from center of pixel $j$ to the tag and $d_{kj,2}$ and is the distance from center of pixel $j$ to the reader antenna corresponding to link $k$.
$\Theta$ is a parameter describing ellipsoid's width. 
Using the general equation for calculating the first Fresnel zone radius at any point in the middle of the link, we chose $\Theta$ to be as follows: 
\begin{equation*}
\Theta = \Theta_0\times\sqrt{\frac{\lambda_{avg}\cdot d_{kj,1}\cdot d_{kj,2}}{d_{kj,1}+d_{kj,2}}}
\label{eqn:solution8}
\end{equation*}
where we chose $\Theta_0$ is an environment and RFID infrastructure dependent parameter, which we tune to achieve reasonable imaging results.
Finally, TagSee employs regularization techniques mentioned in \cite{wilson2010radio,wagner2012passive} to construct image by determining attenuation values per \textit{voxel} (\ie $\mathbf{\hat{x}}$) as $\mathbf{\hat{x}} = \mathbf{P}\times \mathbf{y_{rss}} = (\mathbf{W}^T\mathbf{W} + \sigma_N\mathbf{C_x}^{-1} + \alpha(\mathbf{D}_X^T\mathbf{D}_X + \mathbf{D}_Y^T\mathbf{D}_Y))^{-1}\mathbf{W}^T \mathbf{y_{rss}}$, 
where $\mathbf{D_x}$ and $\mathbf{D_y}$ are differential operators along X and Y directions respectively, catering for the ``spread'' of the impact of attenuations in RSS values along these axes.
We tune $\alpha = 15$ in our current implementation, for reasonable imaging results . 
$\mathbf{C_x}^{-1}$ is another prior term, which controls the spatial correlation of the impact of RSS attenuations across neighboring pixels of the image. 
Although we do not know $\mathbf{\hat{x}}$ in advance, we approximate $\mathbf{C_x}$ based on a exponential spatial attenuation model \cite{agrawal2009correlated} \cite{ripley2005spatial}, as $\mathbf{C_x} = \frac{\sigma_x}{\delta} \exp^{-\mathbf{D_p}/\delta}$,
where $\mathbf{D_p}$ is a square form distance matrix, containing distance between each pair of pixels along the imaginary image plane, and $\sigma_x$ is prior pixel variance due to human motion.
$\sigma_N$ is the prior variance of noise in pixel values \ie when there is no human motion. 
%
%
We approximate $\sigma_N$ as function $c_n*\sigma_{\mathbf{y}}$ of variance in RSS values during the calibration phase. We approximate pixel variance $\sigma_x$ as function $c_x*\sigma_{\mathbf{y}}$ of variance in RSS values during the monitoring phase. We tune the values of $c_n$, $c_x$ and $\delta$ for the best results.

\smallskip\noindent\textbf{Imaging:} For imaging, TagSee erects multiple image planes along Z-axis and then takes an average of the images corresponding to all planes.
We tune the locations of these image planes for best results.
Finally, TagSee combines the images obtained from all antennas by taking their average.
\textit{While evaluating TagSee's performance, we compare the imaging performance of this analytical measurement model based imaging approach with our DNN based approach. }

Next, we propose our DNN based imaging approach, which not only gets rid of manual tuning of parameters in our analytical imaging approach, but also enables multi-person imaging that is required for monitoring browsing behavior of multiple customers towards different item categories simultaneously.

\presec
\section{Deep Learning based RFID Imaging} \label{sec:neuralnet}
\postsub
In our deep learning based RFID imaging approach, we aim to solve the linear regression problem posed in the equation $\mathbf{y_{rss}} = \mathbf{W} \mathbf{x} + \mathbf{n}$ by modeling it as a \textit{deep regression} problem.
Deep regression techniques have been shown to yield state-of-the-art results without having to resort to more complex and ad-hoc regression models \cite{lathuiliere2019comprehensive}.
To achieve this, we model $\mathbf{x} = \mathbf{P}\times\mathbf{y_{rss}}$ as a DNN, where $\mathbf{y_{rss}}$ corresponds to the input of first layer, $\mathbf{x}$ corresponds to the output of last layer and the image construction matrix $\mathbf{P}$ corresponds to the combination of all the layers in between input and output.
The intution is that if we train our system with approximate images of how a human obstacle should look like while he browses a item category, it can automatically learn the underlying relationship between those images and the RFID channel dynamics observed during that browsing activity, which is otherwise difficult to model through geometrical or measurement models based approaches.

%
%

\smallskip\noindent\textbf{Choice of Layers in Neural Network.} We chose the DNN layers such that the weights and biases learned at different layers of the network can mimic the impact of human obstructions on the RFID signals.
Based on the solution to our analytical imaging approach, we can see that image construction matrix $\mathbf{P}=(\mathbf{W}^T\mathbf{W} + \sigma_N\mathbf{C_x}^{-1} + \alpha(\mathbf{D}_X^T\mathbf{D}_X + \mathbf{D}_Y^T\mathbf{D}_Y))^{-1}\mathbf{W}^T$ captures linearities as well as non-linearities related to the impact of human obstructions in terms of attenuations introduced in the RFID channel.
The non-linearities are introduced by terms in weight matrix $\mathbf{W}$, which models the exponential attenuation of RFID signals with distance, and the terms in correlation matrix $\mathbf{C_x}$, which models how the impact of attenuations due to human obstructions decays spatially along the 2D image plane.
The linearities are introducted by the terms corresponding to linear differential operators $\mathbf{D_x}$ and $\mathbf{D_y}$, as well as, due to the inherent linear nature of our imaging problem, \ie all different matrices are connected through basic linear operations such as multiplication, addition, and inverse.
For DNN based imaging, we design the input RSS difference vectors $\mathbf{y_{rss}}$ and the output image vectors $\hat{\mathbf{x}}$ to be normalized vectors, containing values between 0 and 1.
We normalize the input $\mathbf{y_{rss}}$ vectors by dividing with the maximum observed RSS difference, which we empirically estimate for the system.
Moreover, the image construction matrix $\mathbf{P}$ can consist of both positive and negative values, due to the inverse operations involved in it.
%
%

Based on the aforementioned nature of underlying parameters, we chose the first two layers of the DNN to be \textit{rectified-linear} layers (ReLU), next two layers to be \textit{tanh} layers and the last two layers to be \textit{sigmoid} layers.
The output of a ReLU layer is always non-negative real numbers, the output of a \textit{tanh} layer remains between -1 and 1, and the output of a \textit{sigmoid} layer always stays between 0 and 1.
We propose to generalize the dimensions of aforementioned layers as \{$K \times \frac{3}{2}K$\}, \{$\frac{3}{2}K \times 3K$\}, \{$3K \times 3K$\}, \{$3K \times 2K$\}, \{$2K \times 2K$\} and \{$2K \times p_xp_y(k_x-1)(k_y-1)$\}, respectively, where $K = 116$, $k_x = 29$, and $k_y = 4$ in our current implementation of TagSee.
%
%
The number of DNN layers, as well as the dimension of each DNN layer, can be tuned empirically to achieve lowest \textit{cross-validation} errors, and based on the intuition that the amount of information is limited by the number of RFID tags, \ie any constructed image is basically an extrapolation of the impact of RSS variations observed for $K$ tags.
For example, in our proposed DNN architecture, we keep the dimensions of all layers to be within 3 times the number of tags being used for imaging.
Finally, to prevent DNN from over-fitting, we use L2 regularization in combination with \textit{dropout} \cite{SrivastavaDropout}, where the dropout rate ($\ie$ the probability to retain a DNN unit during training) is 0.7.
Figure \ref{fig:neuralnetwork} shows the visual representation of the DNN that we design for our current implementation.

\smallskip\noindent\textbf{Training Requirements.}
We separately ask a couple of volunteers to browse each different item category by standing in front of them, for approximately $30-60~secs$.
\textit{We do not constrain natural human motion during this training phase, \ie, the individuals browsing different item categories are allowed to move to-and-fro and sideways, and browse items in a natural manner.}
%
%
TagSee uses the variations observed in RSS values of the deployed tags during this training phase as inputs to its DNN, after normalization.
%
%
For each browsing activity during the training phase, we also generate approximate normalized images of human obstructions, which TagSee uses as the training ``labels'' or regression outputs to its DNN.
In this work, we approximate the images of human subjects as ellipses, where the width and height of those ellipses is chosen as average width and height of the human subjects used for training TagSee's DNN.
The height of the ellipses is limitted by the size and location of the deployed mesh of tags.
%
%
For robustness, we design TagSee's DNN classification module as an ensemble of $D$ single DNN classifiers. The final output is median of the outputs obtained from all the $D$ single classifiers.
%

%

\begin{figure*} [htbp]
	\centering
	\captionsetup{justification=centering}
	\subfigure[ \noindent{Spatial moving window of impact width $k_{cw} = 6$}]{
		\label{fig:multihumanimaging1a}
		\includegraphics[width=0.36\textwidth]{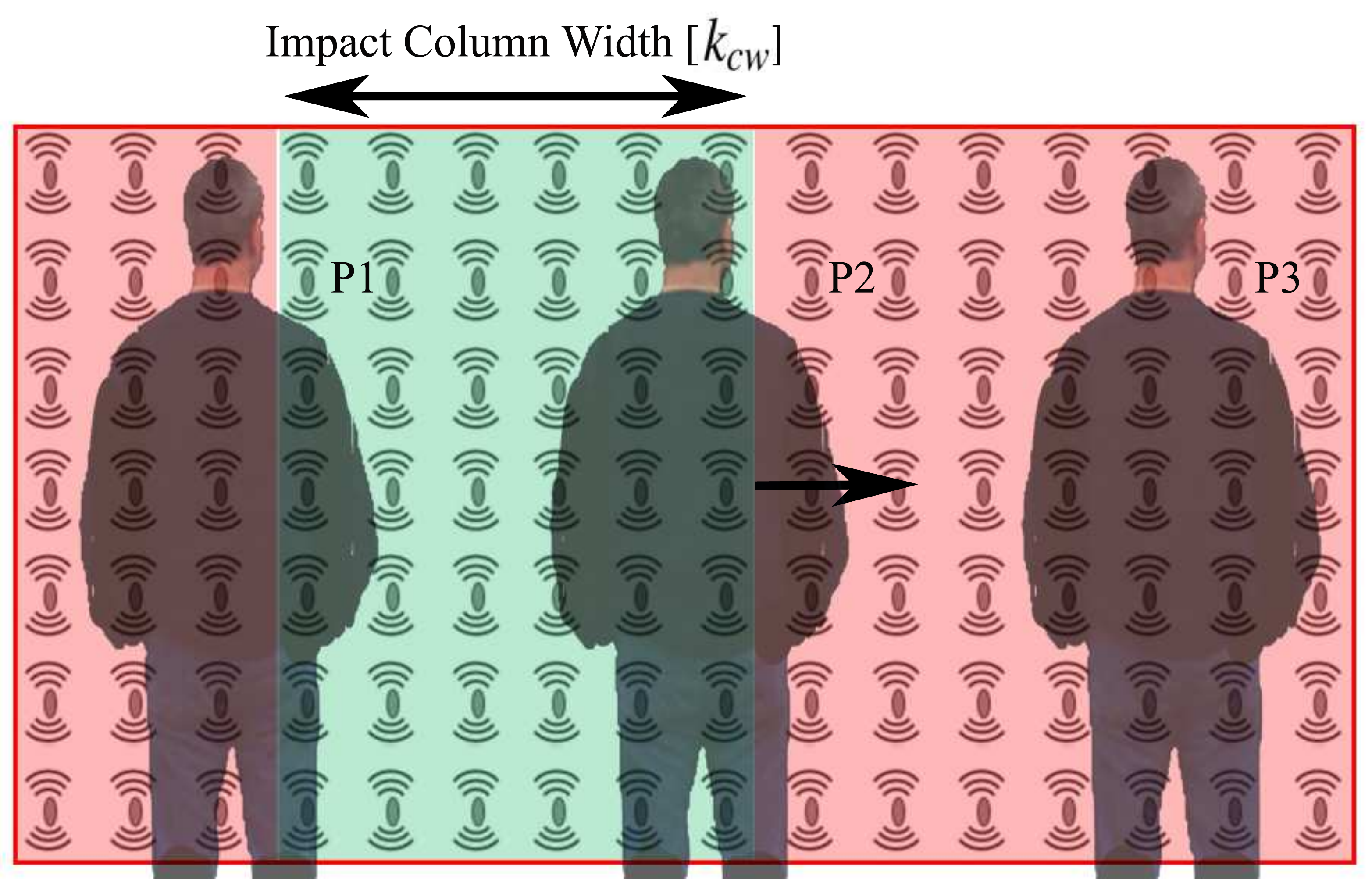}%
	}
	\hspace{0.2in}
	\subfigure[\noindent{Distribution of $\mathbf{y_{rss}}$ during \textit{calibration mode}}]{
		\label{fig:multihumanimaging1b}
		\includegraphics[width=0.392\textwidth]{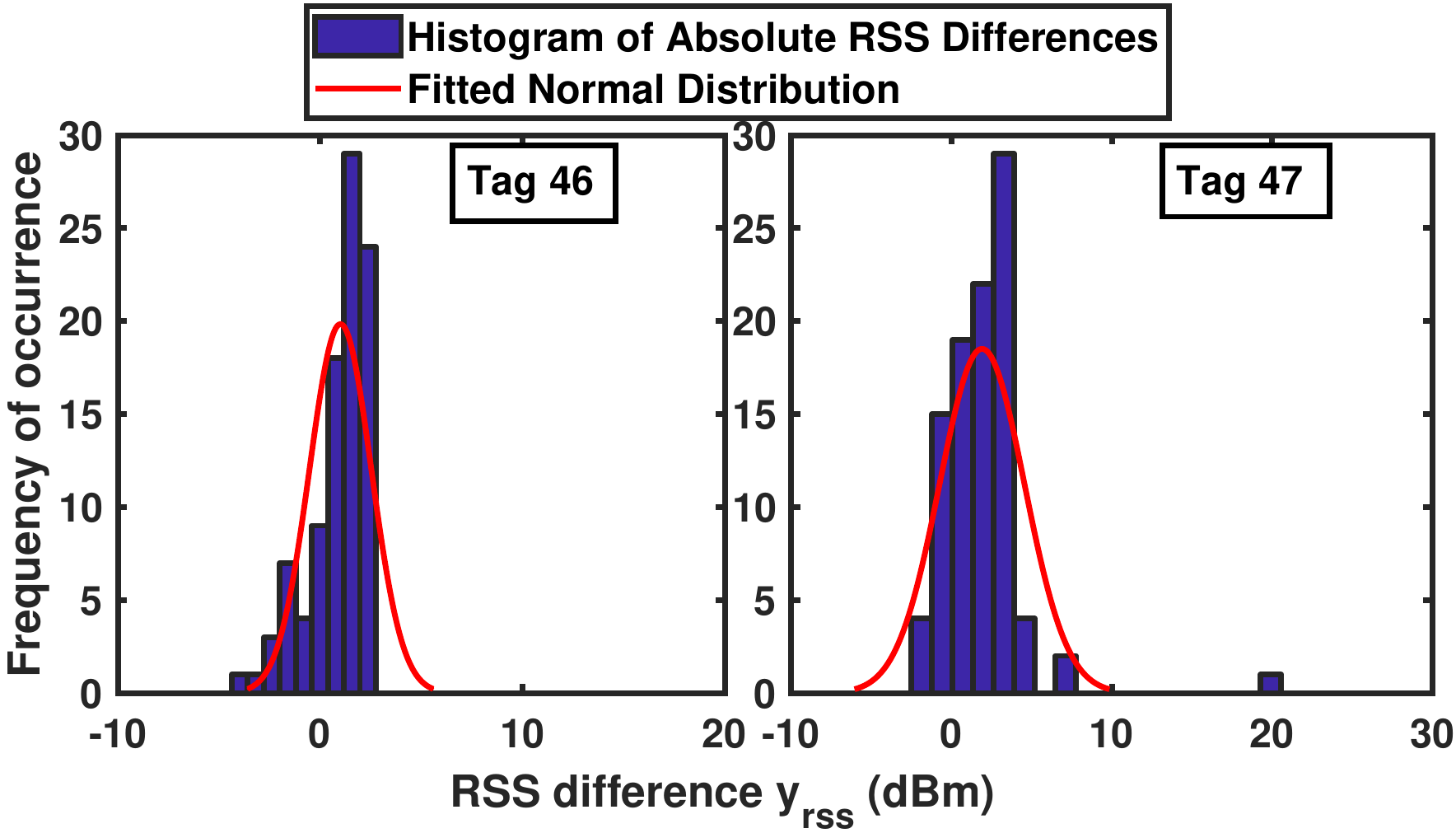}%
	}
	\vspace{-0.1in}
	\caption{TagSee's spatial moving window based approach for multi-person imaging} 
	\label{fig:multihumanimaging1}
	\vspace{-0.15in}
\end{figure*}
\begin{figure*}[htbp]
	\centering
	\captionsetup{justification=centering}	\includegraphics[width=2.07\columnwidth]{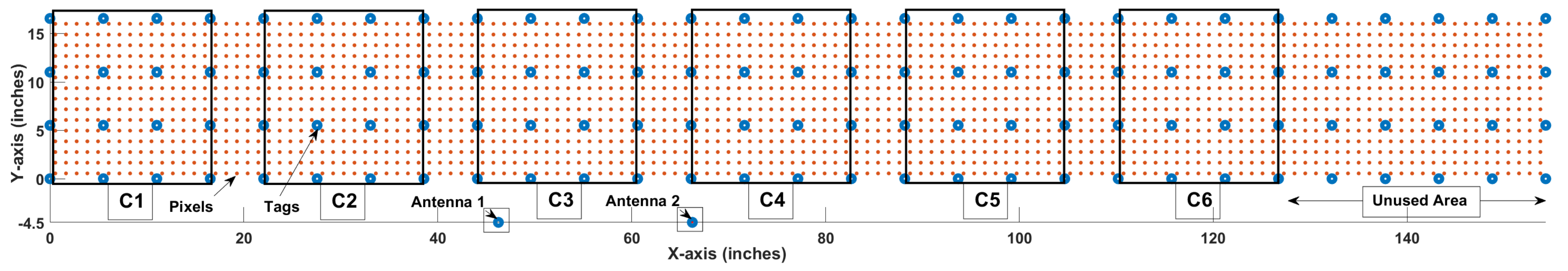}
	\vspace{-0.23in}
	\caption{Detailed experimental setup} 
	\label{fig:setup}
	\vspace{-0.18in}
\end{figure*}

\presec
\section{Multi-Person RFID Imaging} \label{sec:multihuman}
\postsec
%
%
%
To develop our multi-person imaging technique, TagSee leverages the intuition that a human subject will only impact $k_{cw}$ columns of tags along X-axis, for any deployment of a mesh of tags.
%
%
%
%
%
Based on this intuition, TagSee creates multiple new vectors \{$\mathbf{y_{rss}^i}$\} from each new RSS difference vector $\mathbf{y_{rss}}$ corresponding to an antenna, by moving a window of size $k_{cw}$ columns over the spatial distribution of deployed tags, as shown in figure \ref{fig:multihumanimaging1a}.
For each window $i$, TagSee only copies those values from $\mathbf{y_{rss}}$ to $\mathbf{y_{rss}^i}$ which correspond to the tags contained inside that window,
and replaces the values corresponding to remaining tags with two times the standard deviation values ($2\cdot\sigma_{k,a,rss}$) of Gaussian distributions $\mathcal{N}(\mu_{k,a,rss}, \sigma_{k,a,rss}^2)$, which model the RSS difference values observed for those tags, respectively, when there are no obstructions around.
This avoids spurious blobs during image construction.
%
Figure \ref{fig:multihumanimaging1b} shows the distribution of the RSS difference values, obtained for two closely spaced tags, during calibration mode in one of our experiments, over a period of $\sim2~ mins$.
We can observe that the RSS difference values approximately follow a Gaussian distribution.
The mean and variance values for the aforementioned distributions $\mathcal{N}(\mu_{k,a,rss}, \sigma_{k,a,rss}^2)$ corresponding to each possible tag-antenna pair, are estimated during the calibration phase.

For $k_x$ columns of tags, TagSee first generates $k_x-k_w+1$ new vectors \{$\mathbf{y_{rss}^i}$\}.
Afterwards, it constructs images corresponding to all vectors \{$\mathbf{y_{rss}^i}$\} using the aforementioned DNN based imaging technique, and then merges them after passing through a 2D filter (median and averaging filters), to output the final image.
TagSee applies this multi-person imaging technique for each antenna, separately, and finally combines the images obtained from all antennas through averaging.
%
%
This approach enables multi-person imaging without changing the training requirements of TagSee's imaging technique, \ie our system does not require the DNN to be trained for multi-person scenarios.
The only change required is to train DNN for all possible $\mathbf{y_{rss}^i}$ corresponding to every training sample.
\textit{Note that for any deployment, we train TagSee's DNN with data corresponding to `no obstruction' scenarios as well to avoid detection of spurious blobs during image construction.}
To train the DNN for `no obstruction' scenarios, the input vectors are set to be the $2\cdot\sigma_{k,a,rss}$ values corresponding to each deployed tag and the outputs are set to be zero vectors (i.e. blank images).
%

\smallskip\noindent\textbf{Monitoring Browsing Activity:}
TagSee monitors the customer browsing behavior towards different items in terms of popularity of those items.
%
In monitoring mode, TagSee feeds the final constructed image frames to a \textit{Blob Analysis} module \cite{matlabblobtrack}, which determines the background using a few initial frames, and then outputs the coordinates of bounding boxes and centroids of any human images it detects in foreground of each consecutive frame.
As the boundaries of item categories are known in advance, TagSee determines the \textit{popularity} of each category by checking the proximity of the centroids corresponding to detected blobs in each frame to the centroid of that category.
If the centroid of a blob is within $\eta_2$ \textit{voxels} of the centroid of $j^{th}$ category, TagSee increments the popularity $\mathcal{P}_j$.
$\eta_1$ and $\eta_2$ is dependent upon the density of deployed tags, and can be easily tuned empirically for a certain deployment scenario.
For robust popularity estimates, TagSee maintains a buffer consisting of 5 latest constructed images (which corresponds to a period of $\sim3-4~secs$), takes the median of all those images and applies thresholding (on the scale from 0 to 1, pixel values below 0.1 are set to 0), before calculating $\mathcal{P}_j$'s.
\presec\presec
\section{Implementation \& Evaluation} \label{sec:Implementevaluate}
\postsec
We implement TagSee using COTS UHF Impinj R420 pRFID reader \cite{impinjr420} and SMARTRAC's Dogbone pRFID tags \cite{SMARTRACDOGBONE}, which operates in frequency range 902.75 - 928.25 MHz and is compatible with EPC Global C1G2 \cite{epcglobal2008radio} standard.
We use two circular polarized antennas, which are connected to two of the four antenna ports of R420 reader.
As reader interrogates the deployed \textit{static} tags, the information containing IDs, time stamps, channel frequencies, reader antenna IDs, RSS values and phase values corresponding to the tags read in each cycle are sent through Ethernet to a laptop running TagSee.
We develop our RFID data collection module by bulding upon the JAVA based Impinj Octane SDK \cite{octanesdk}.
%
%
For any deployment scenario, TagSee first runs in \textit{calibration mode}, for approximately $2~mins$, to determine the background values of RSS and phase for all deployed tags. 
%
%
Afterwards, it turns on its \textit{monitoring mode} to image customers and track popularity of different item categories being browsed by those customers.
%
\begin{figure*} [htbp]
	\centering
	\captionsetup{justification=centering}
	\includegraphics[width=0.93\textwidth]{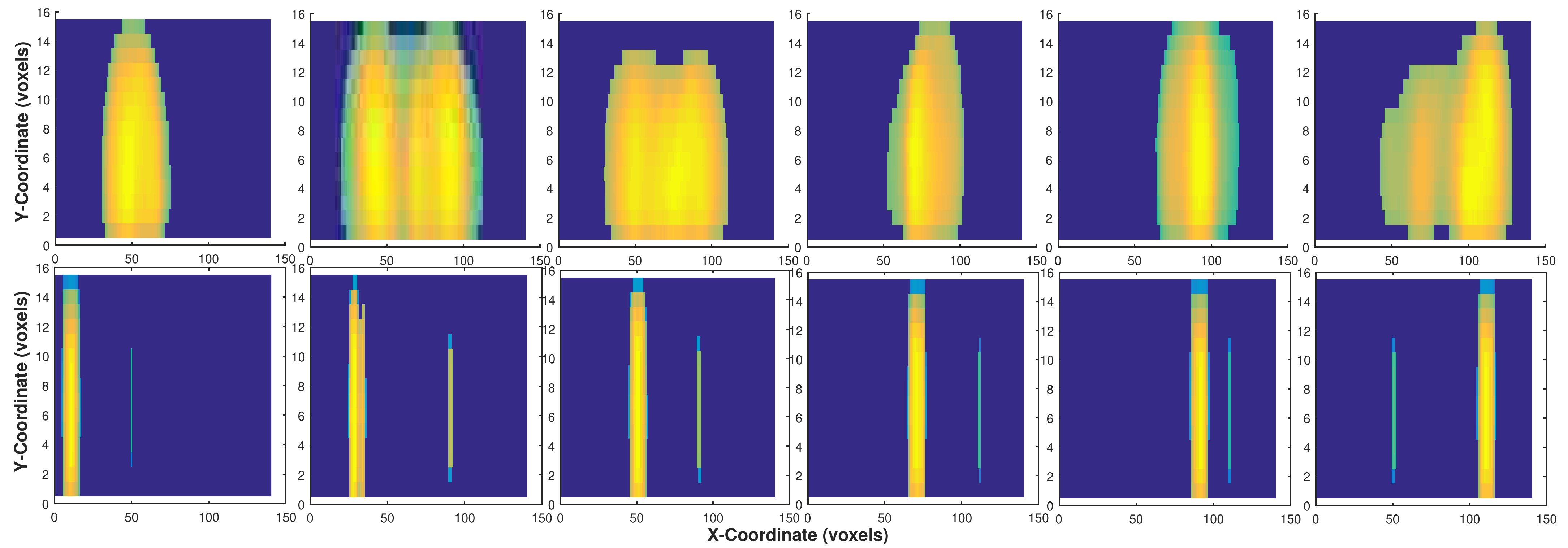}%
		\vspace{-0.05in}
	\caption{Comparison between TagSee's baseline (top) and DNN based (bottom) approaches for single person scenario.} 
	\label{fig:user7testsignleperson}
	\vspace{-0.12in}
\end{figure*}
\begin{figure*} [htbp]
	\centering
	\captionsetup{justification=centering}
	\subfigure[ \noindent{Average error rates for test user 1}]{
		\label{fig:signlepersontraining1}
		\includegraphics[width=0.30\textwidth]{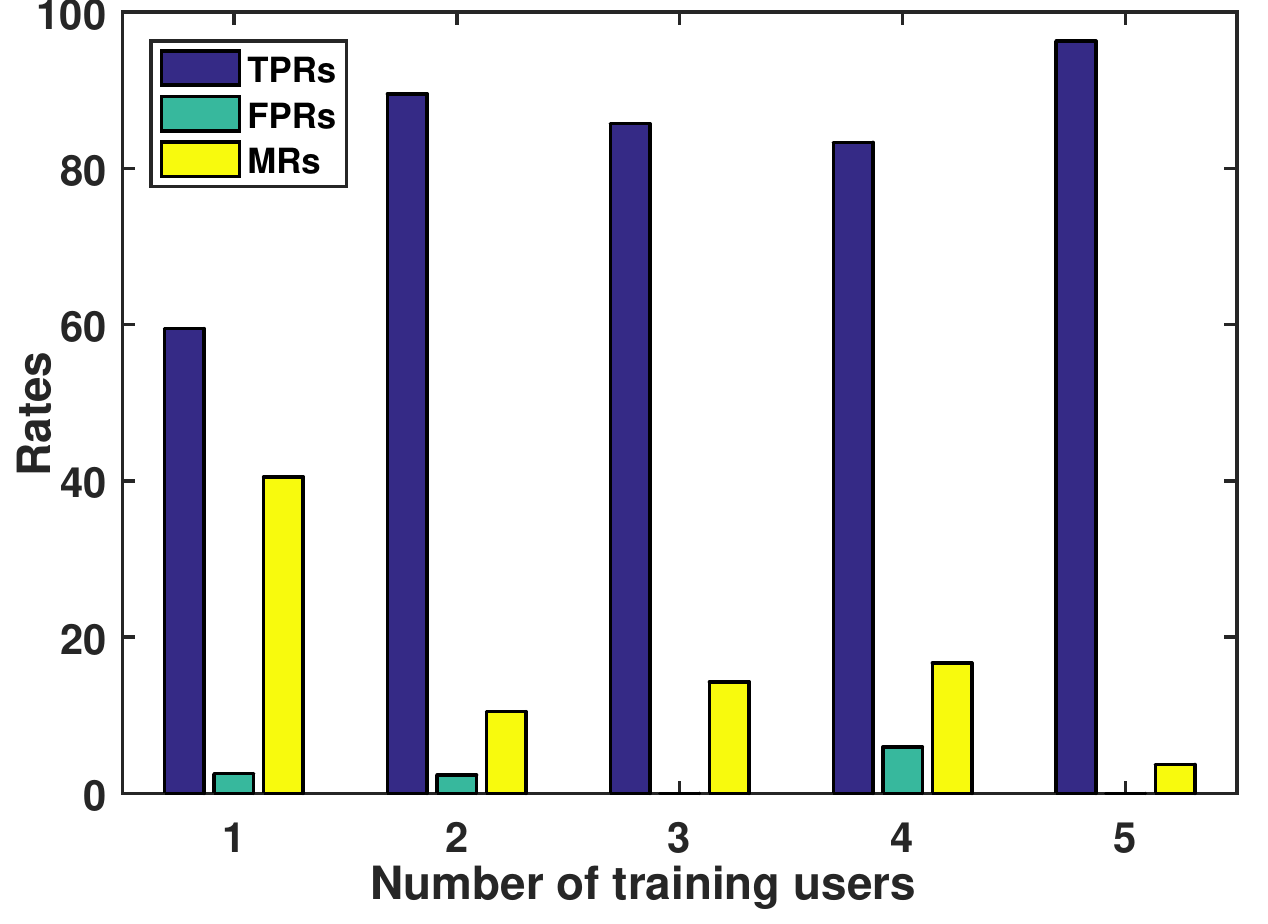}%
	}
	\subfigure[ \noindent{Average error rates for test user 2}]{
		\label{fig:signlepersontraining2}
		\includegraphics[width=0.30\textwidth]{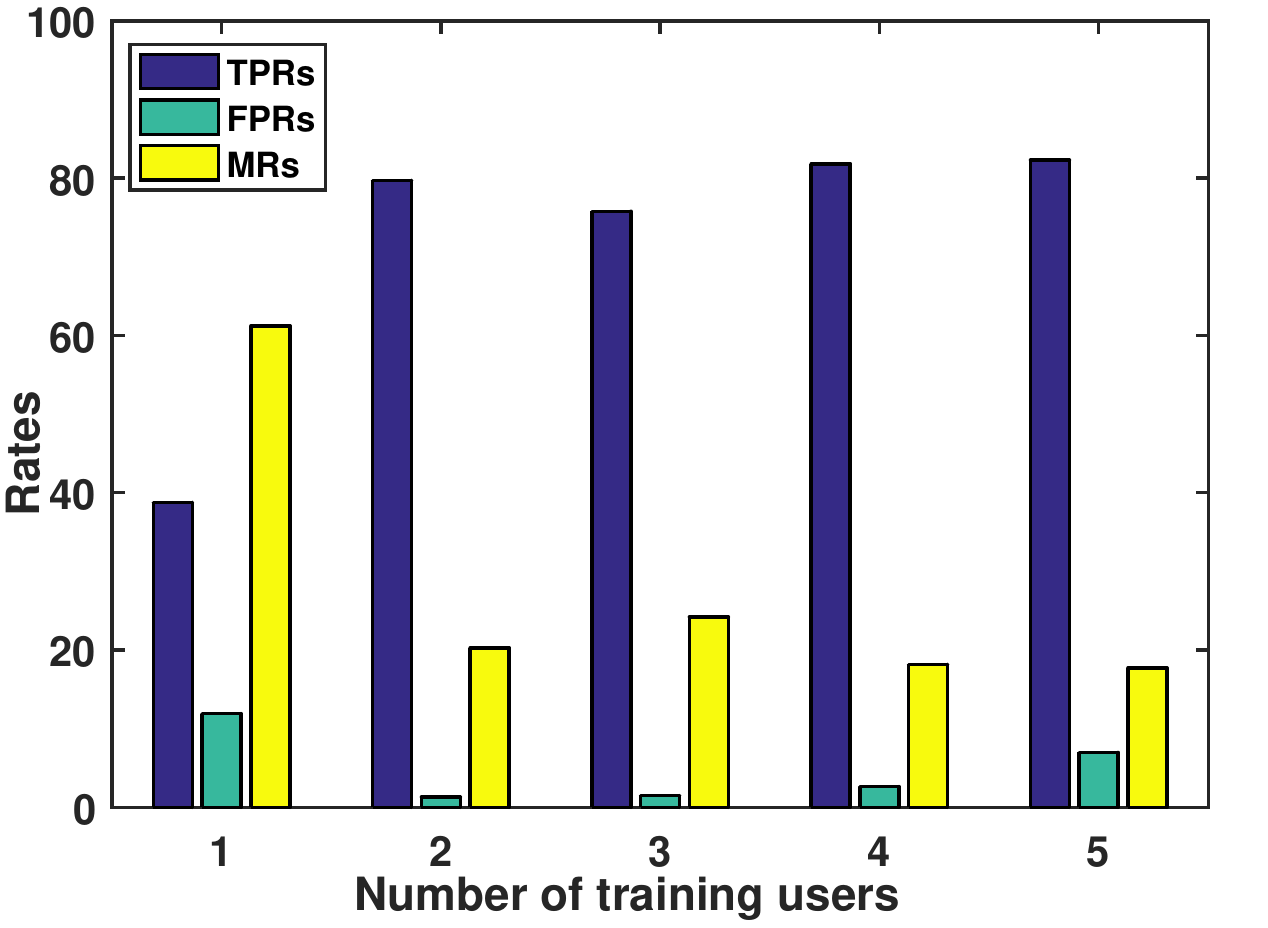}%
	}
	\subfigure[ \noindent{Average error rates for test user 3}]{
		\label{fig:signlepersontraining3}
		\includegraphics[width=0.30\textwidth]{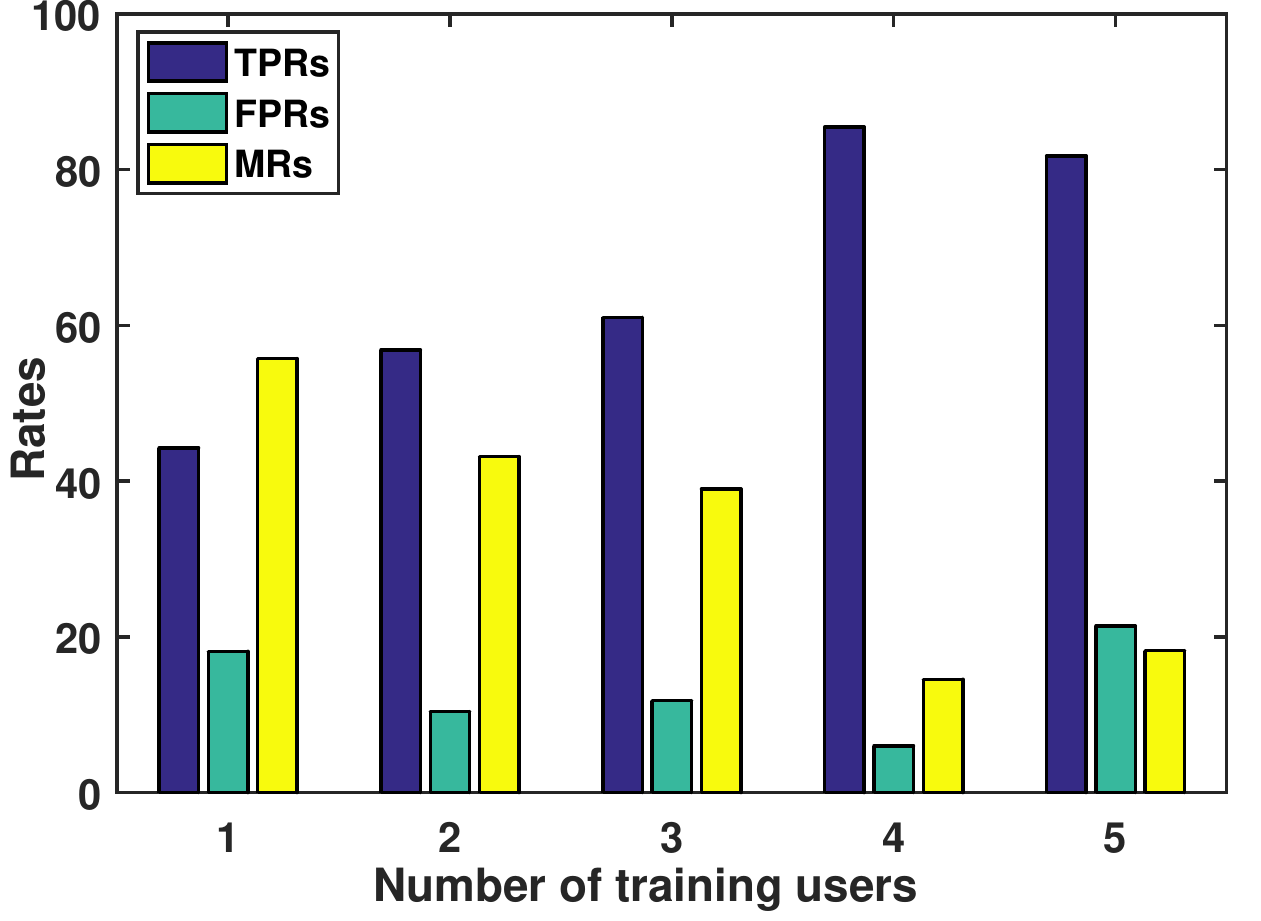}%
	}
    \vspace{-0.08in}
	\caption{Effect of number of training users on TagSee's performance (TPRs, FPRs and MRs) for $k_{cw} = 6$} 
	\label{fig:signlepersontraining}
	\vspace{-0.22in}
\end{figure*}
\begin{figure} [htbp]
	\centering
	\captionsetup{justification=centering}
	\includegraphics[width=0.31\textwidth]{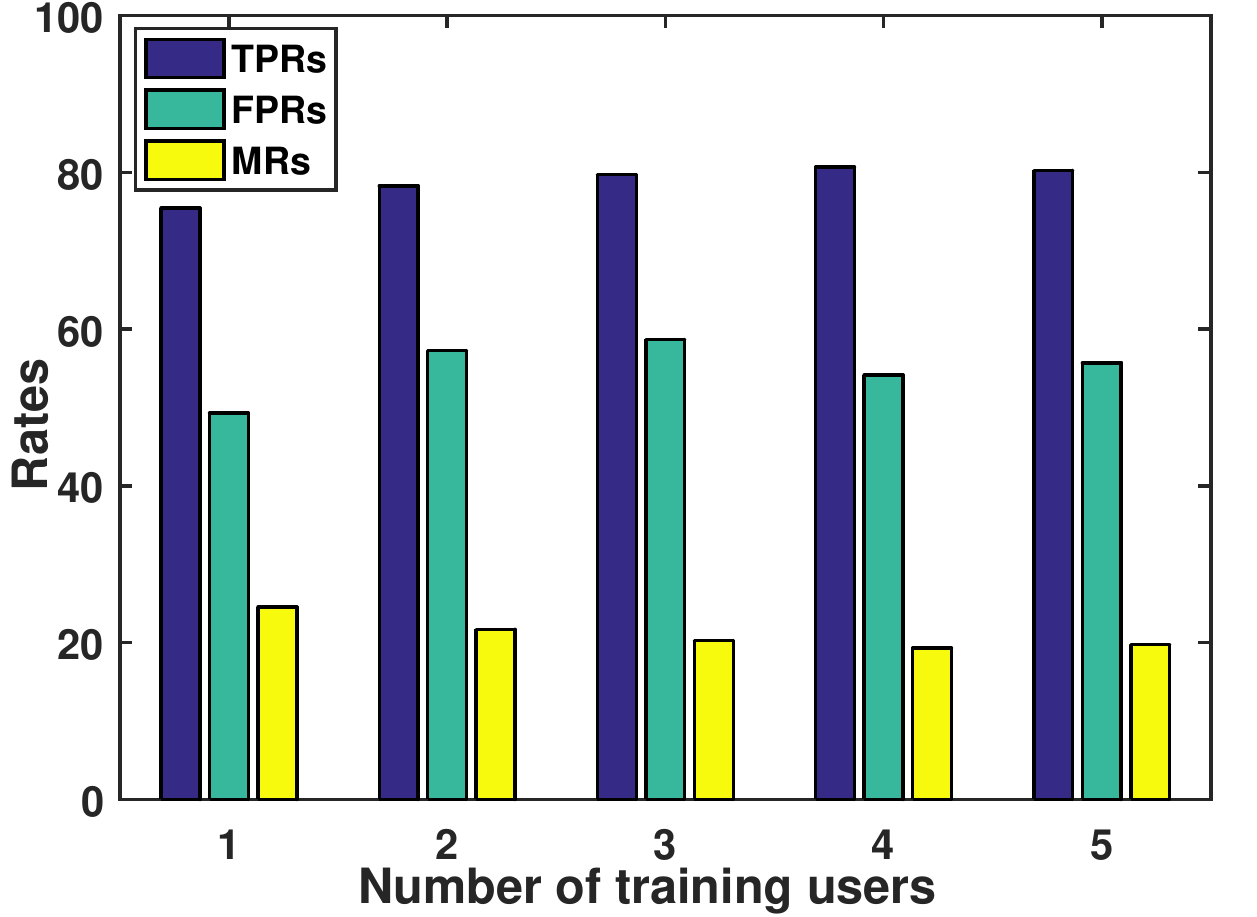}
	\vspace{-0.05in}
	\caption{Performance in single person monitoring scenario using 1 reader antenna only, $k_{cw}=8$} 
	\label{fig:singleperson1antenna}
	\vspace{-0.24in}
\end{figure}
\begin{figure*} [htbp]
	\centering
	\captionsetup{justification=centering}
	\includegraphics[width=0.93\textwidth]{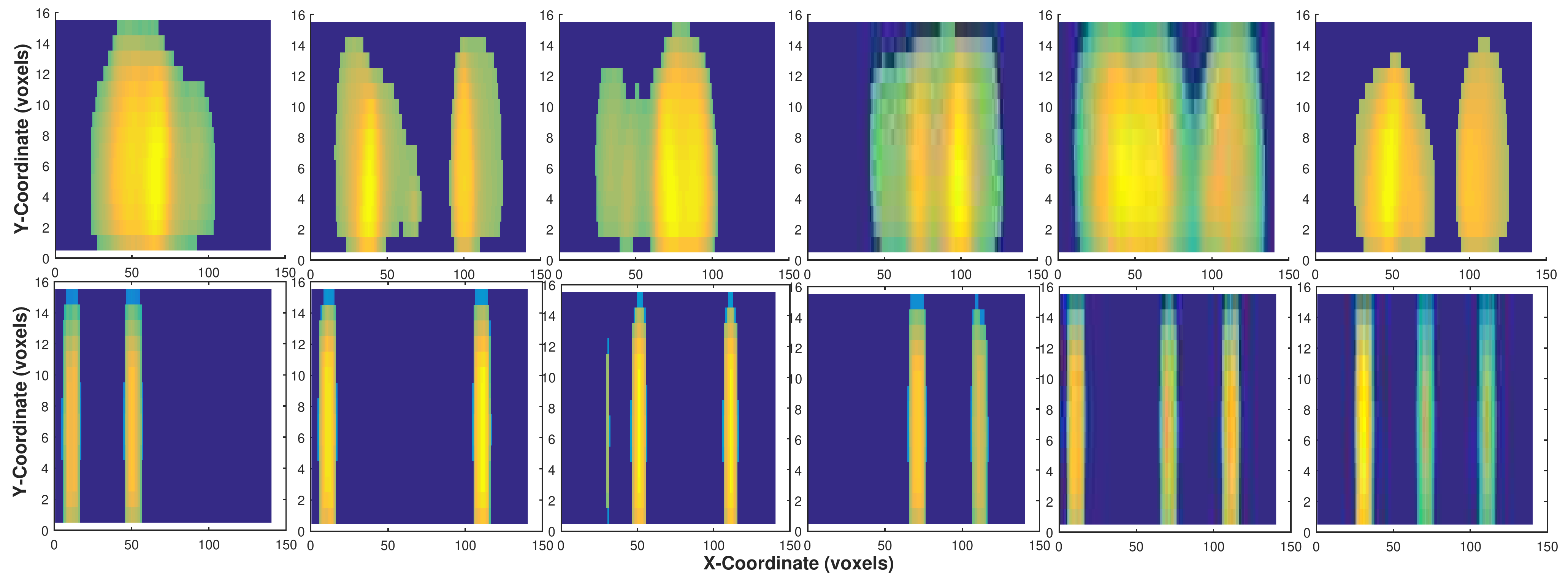}%
	\caption{Comparison between TagSee's baseline (top) and DNN based (bottom) RFID imaging approaches for multi-person scenario. The leftmost 4 images correspond to 2-user scenarios, and the rightmost 2 images correspond to 3-user scanerios} 
	\label{fig:users5_6_7testmultiperson}
	\vspace{-0.2in}
\end{figure*}
\begin{figure*} [htbp]
	\centering
	\captionsetup{justification=centering}
	\subfigure[ \noindent{Overall average error rates for $k_{cw} = 4$}]{
		\label{fig:multipersontrain1}
		\includegraphics[width=0.31\textwidth]{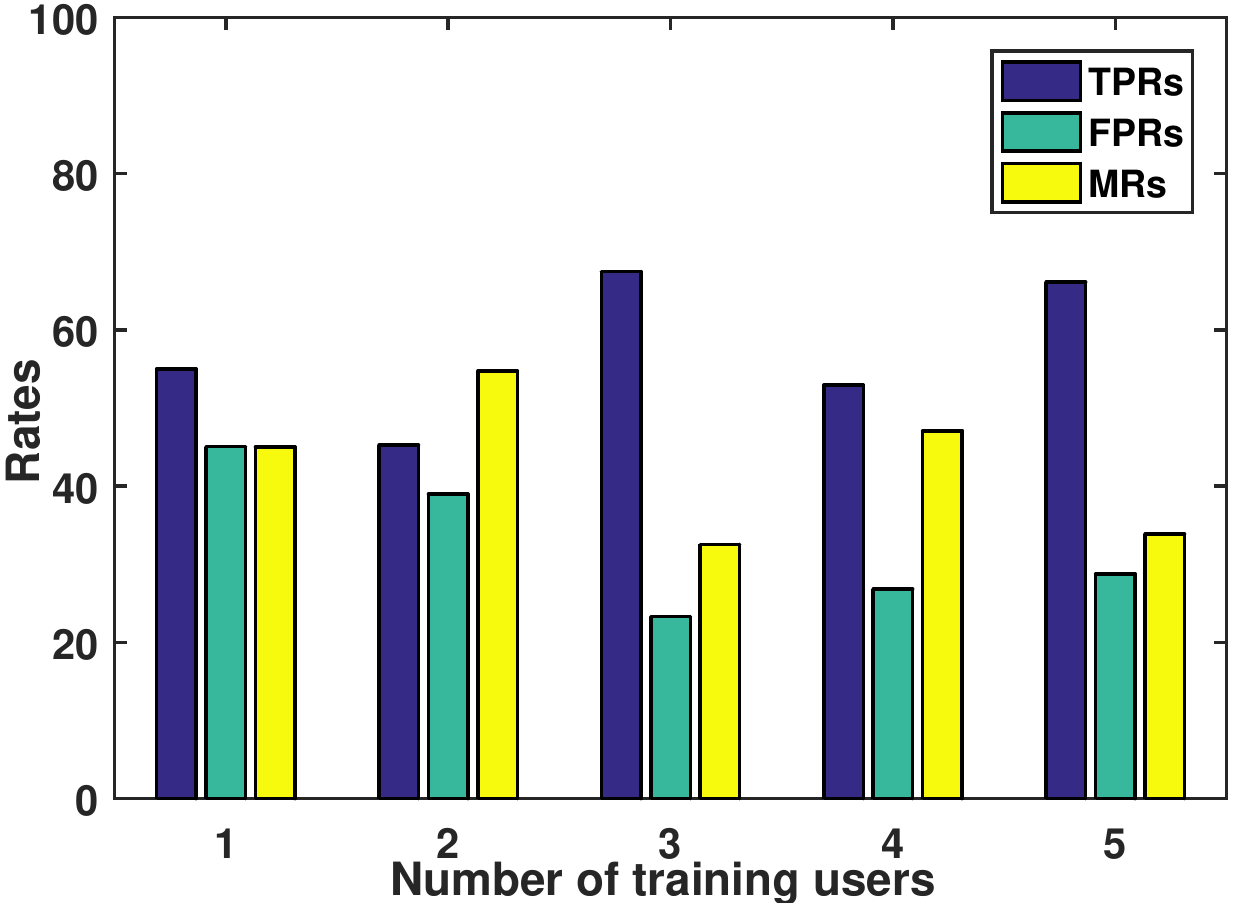}%
	}
	\subfigure[ \noindent{Overall average error rates for $k_{cw} = 6$}]{
		\label{fig:multipersontrain2}
		\includegraphics[width=0.31\textwidth]{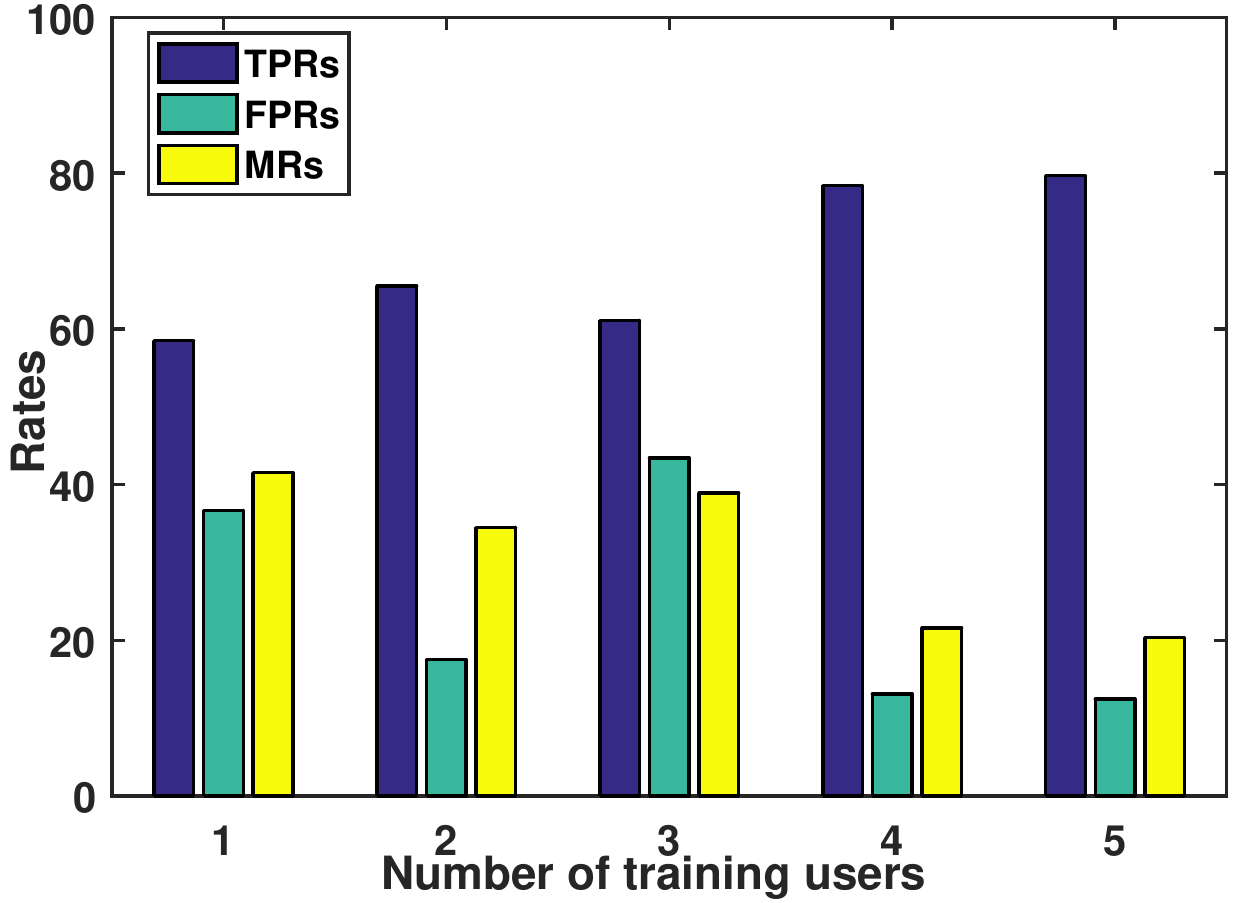}
	}
	\subfigure[ \noindent{Overall average error rates for $k_{cw} = 8$}]{
		\label{fig:multipersontrain3}
		\includegraphics[width=0.31\textwidth]{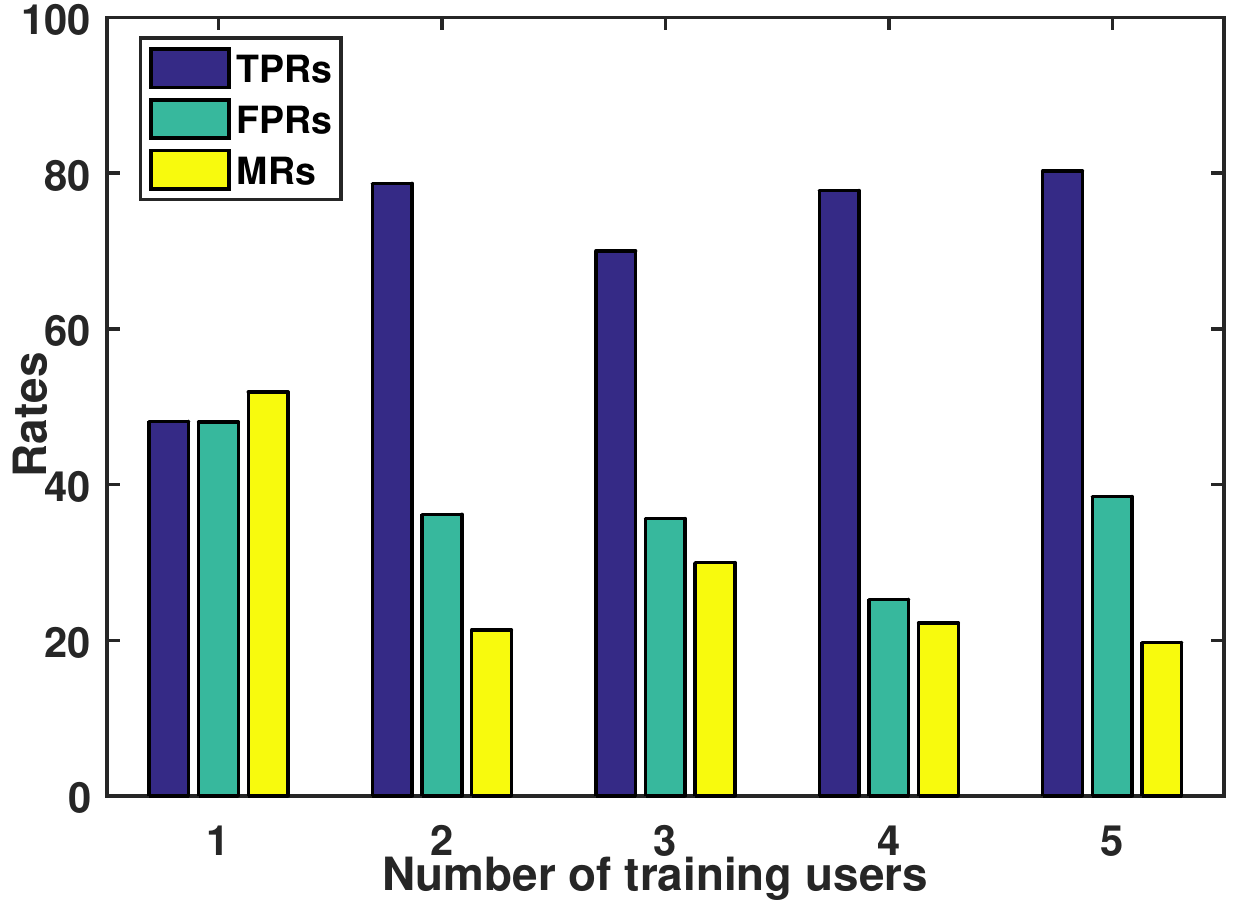}%
	}
	\vspace{-0.1in}
	\caption{Effect of impact width $k_{cw}$ and number of training users on TagSee's performance in 2 person scenarios} 
	\label{fig:multipersontrain}
	\vspace{-0.2in}
\end{figure*}
\begin{figure*} [htbp]
	\centering
	\captionsetup{justification=centering}
	\subfigure[ \noindent{Per scenario error rates for $k_{cw} = 4$}]{
		\label{fig:multipersontrain1_a}
		\includegraphics[width=0.31\textwidth]{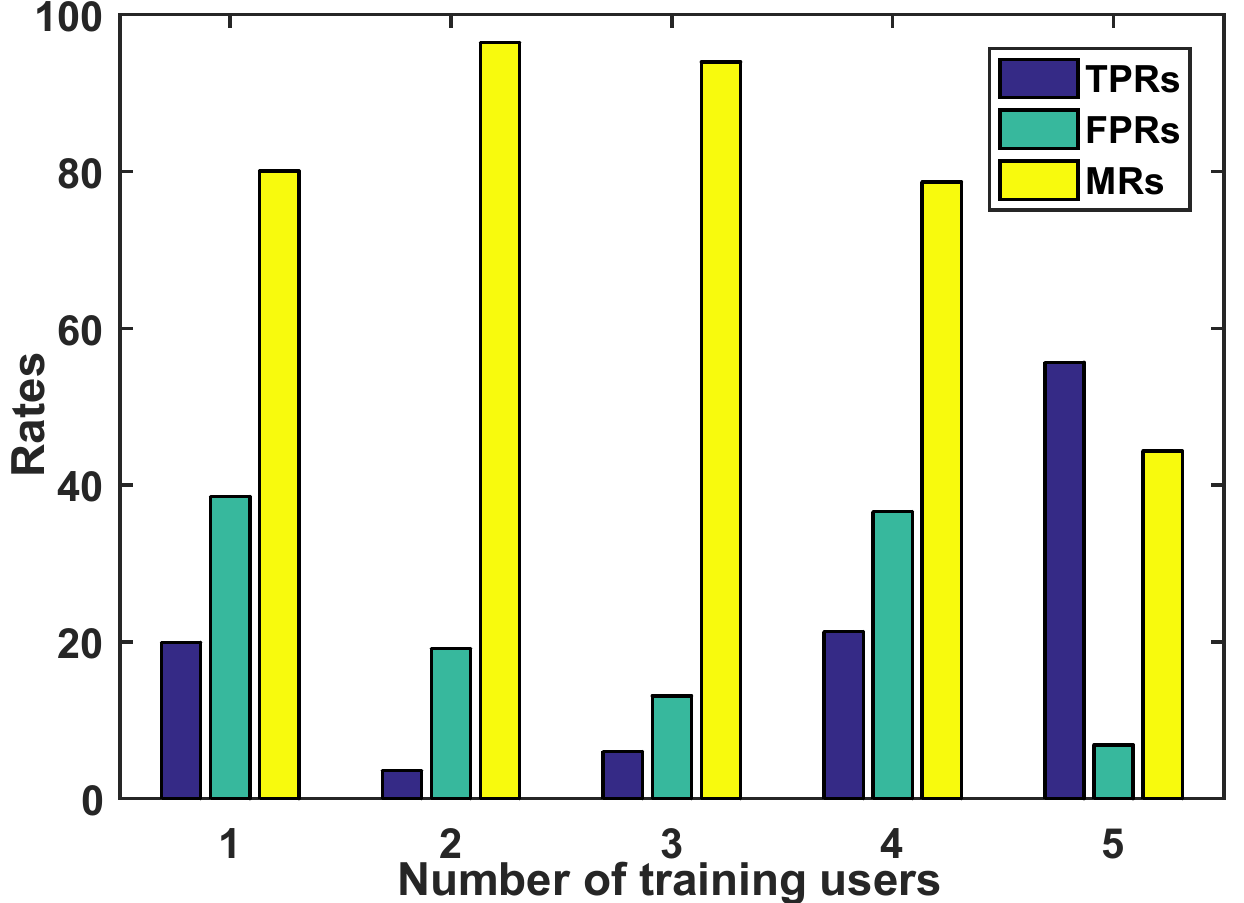}%
	}
	\subfigure[ \noindent{Per scenario error rates for $k_{cw} = 6$}]{
		\label{fig:multipersontrain2_a}
		\includegraphics[width=0.31\textwidth]{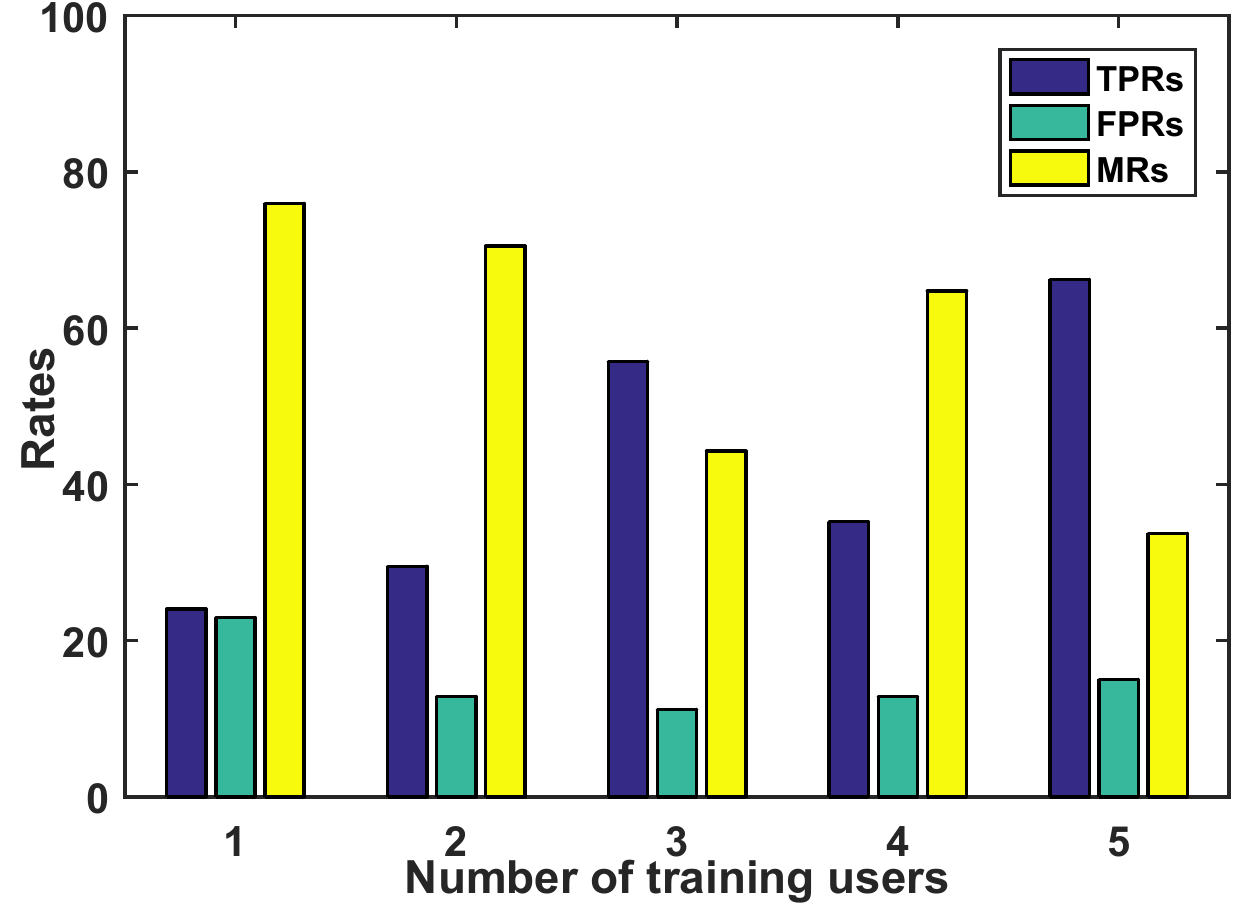}%
	}
	\subfigure[ \noindent{Per scenario error rates for $k_{cw} = 8$}]{
		\label{fig:multipersontrain3_a}
		\includegraphics[width=0.31\textwidth]{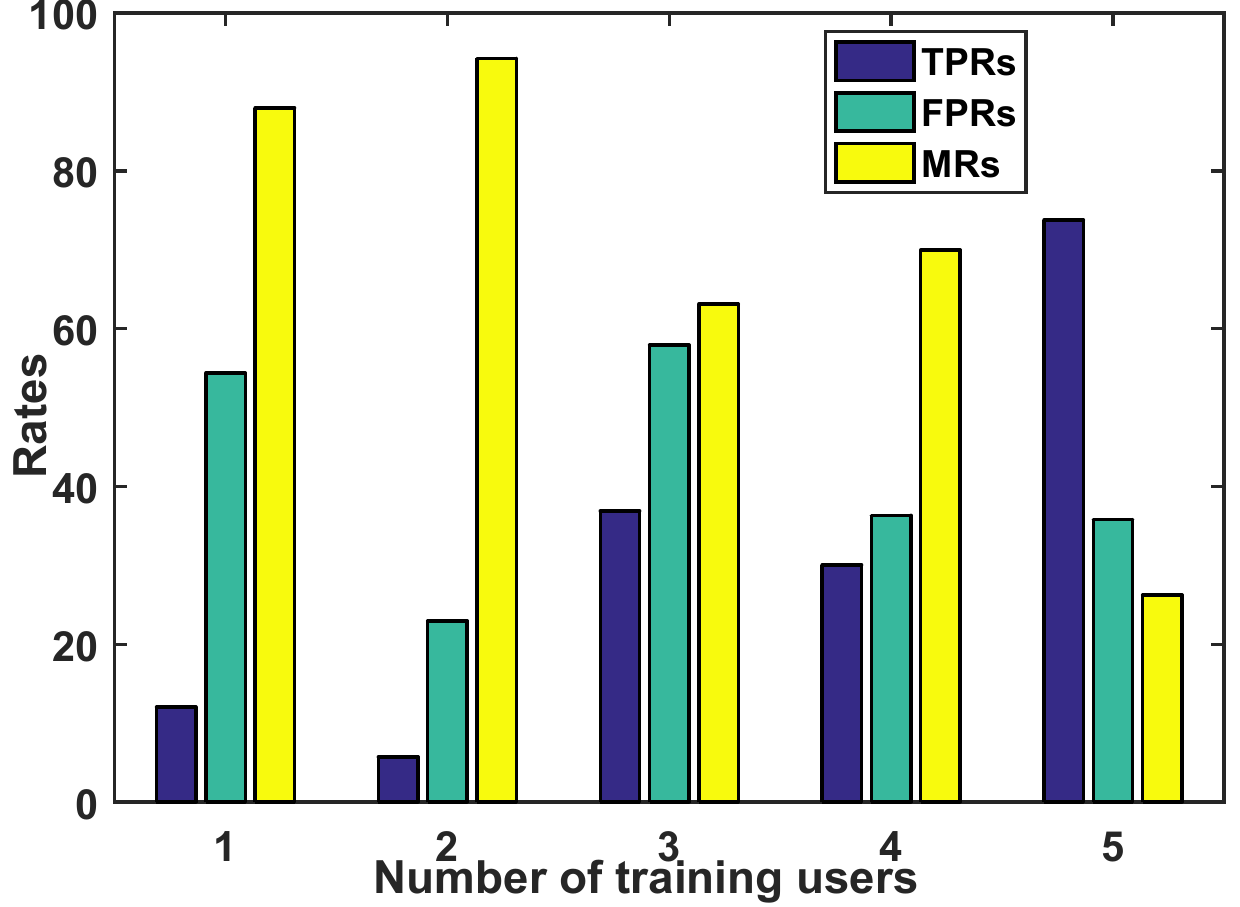}%
	}
	\vspace{-0.1in}
	\caption{Effect of impact width $k_{cw}$ and number of training users on TagSee's performance in 3 person scenarios} 
	\label{fig:multipersontrain_a}
	\vspace{-0.2in}
\end{figure*}
\begin{figure*} [htbp]
	\centering
	\captionsetup{justification=centering}
	\subfigure[ \noindent{Per scenario error rates for $k_{cw} = 4$}]{
		\label{fig:multipersonimpactwidth1}
		\includegraphics[width=0.31\textwidth]{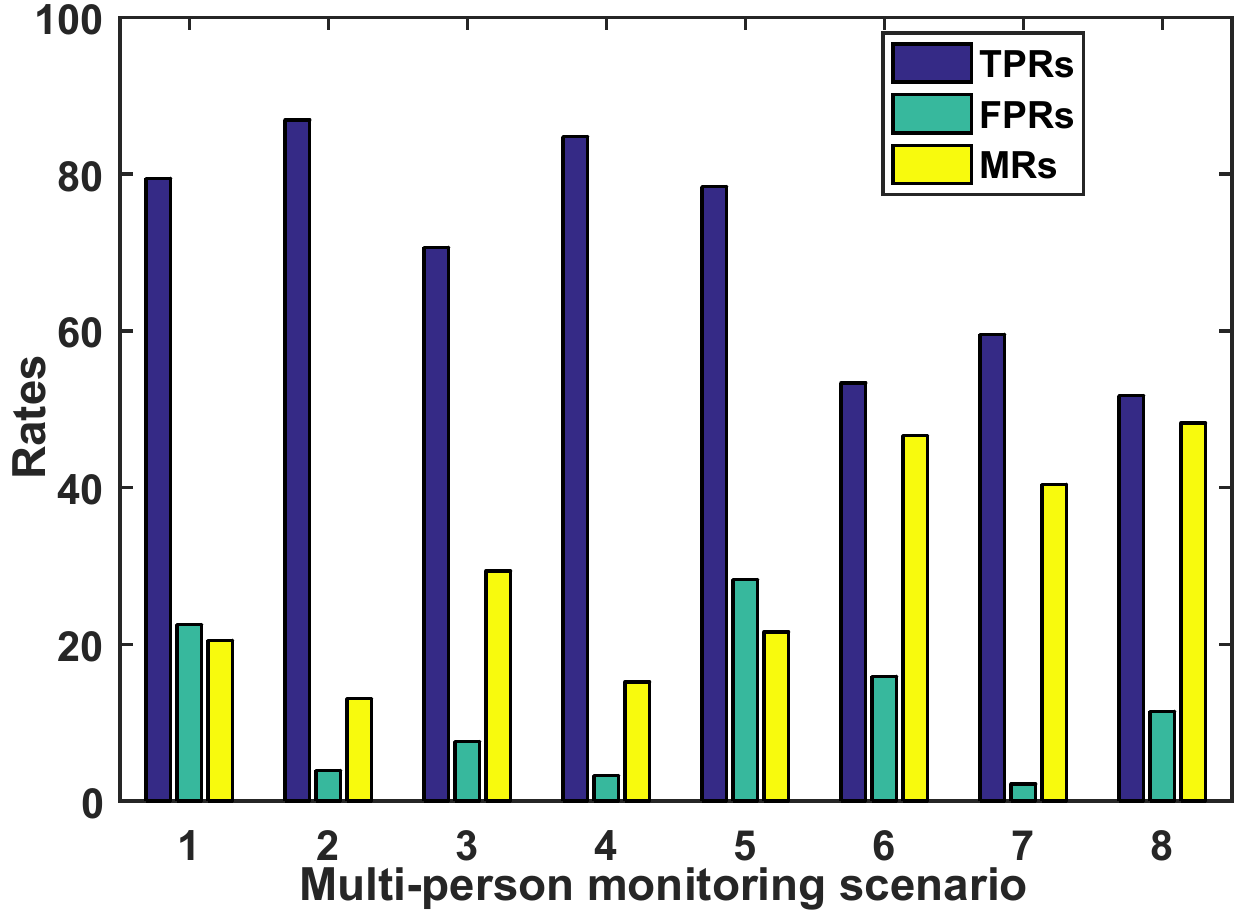}%
	}
	\subfigure[ \noindent{Per scenario error rates for $k_{cw} = 6$}]{
		\label{fig:multipersonimpactwidth2}
		\includegraphics[width=0.31\textwidth]{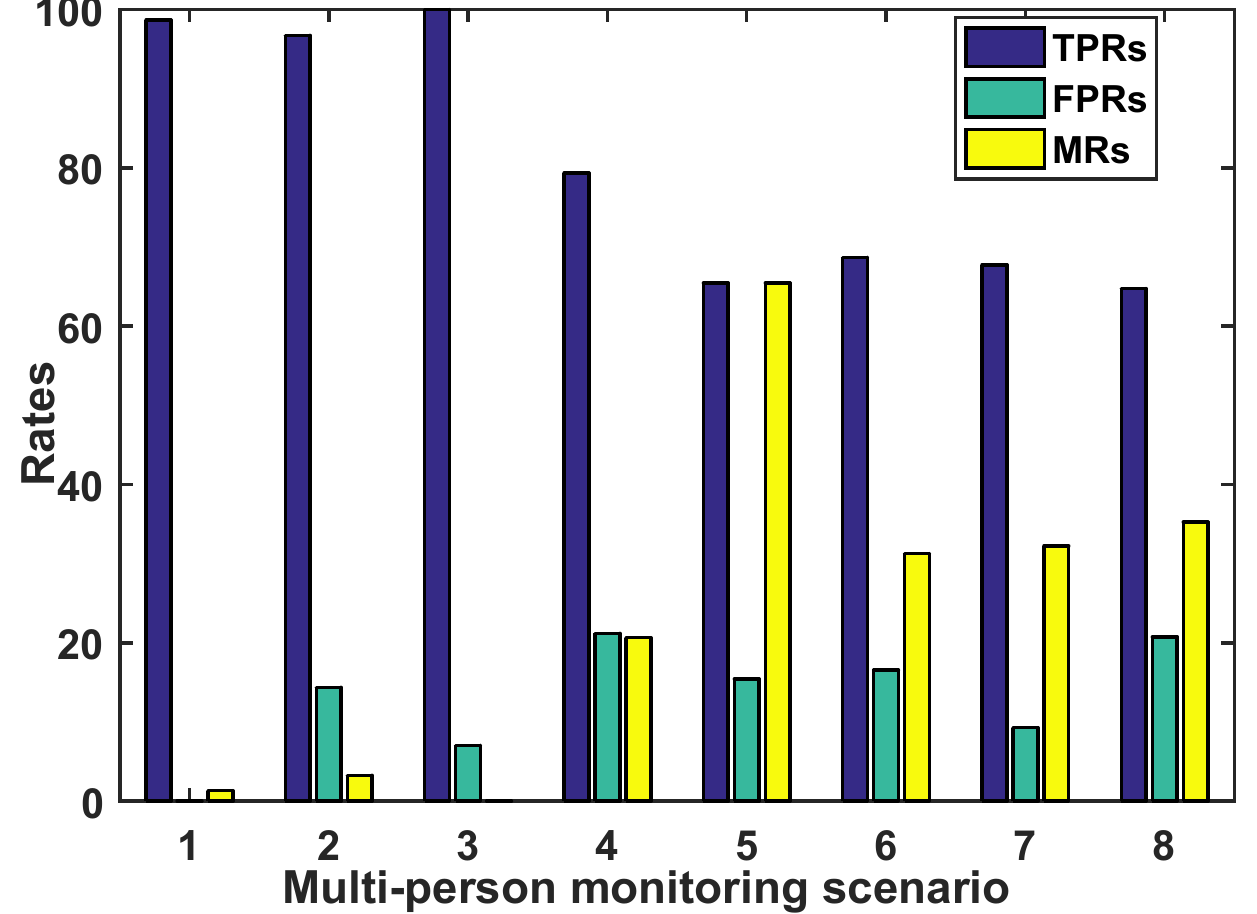}%
	}
	\subfigure[ \noindent{Per scenario error rates for $k_{cw} = 8$}]{
		\label{fig:multipersonimpactwidth3}
		\includegraphics[width=0.31\textwidth]{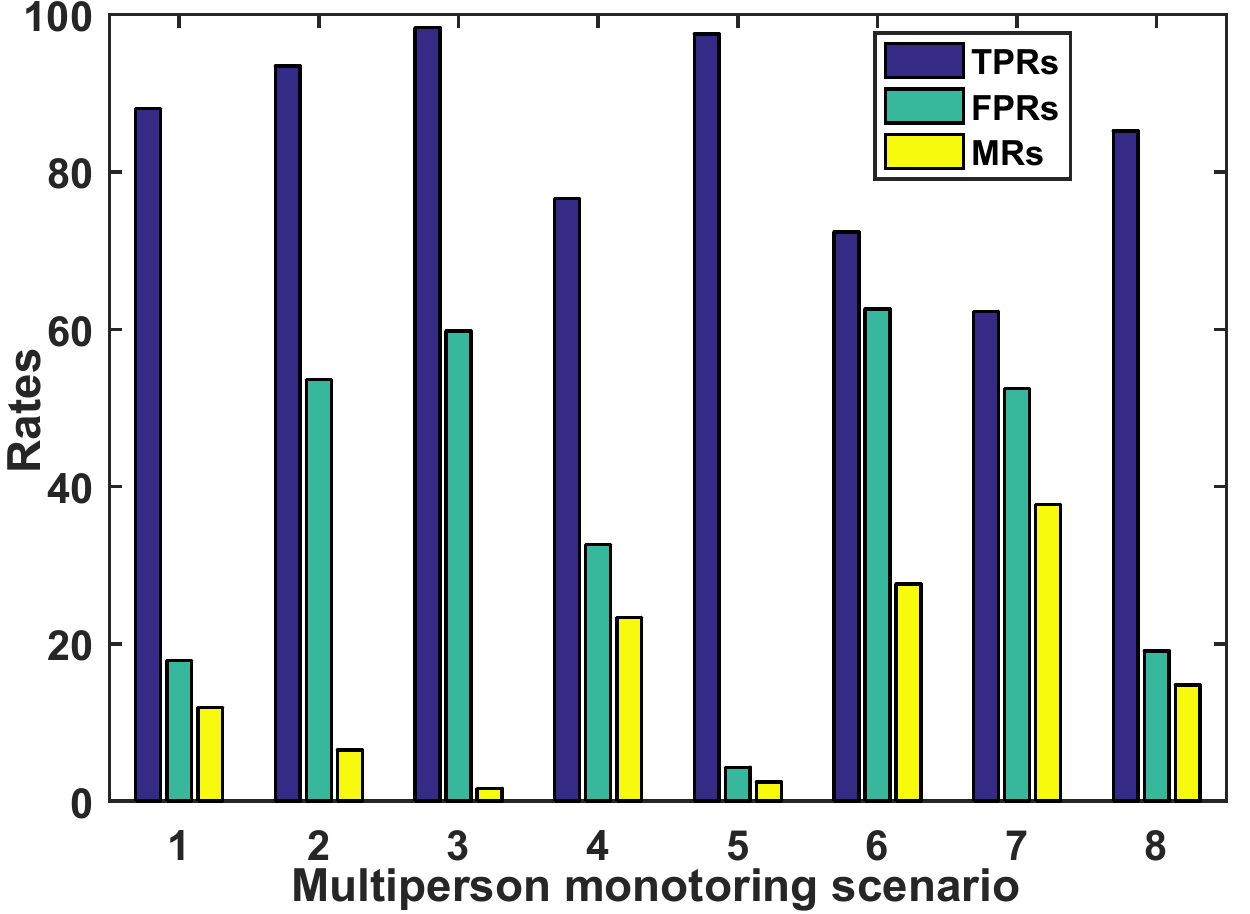}%
	}
	\vspace{-0.1in}
	\caption{Effect of impact width $k_{cw}$ on TagSee's performance (TPRs, FPRs and MRs) for 8 different multi-person scenarios, i.e. item category sets \{1,3\}, \{1,4\}, \{1,5\}, \{1,6\}, \{3,6\}, \{4,6\}, \{1,4,6\}, \{2,4,6\}, 5 training users used} 
	\label{fig:multipersonimpactwidth}
\end{figure*}

\presub\presub
\subsection{Evaluation Methodology}
\subsubsection{Experimental Setup}
Fig. \ref{fig:setup} shows detailed experimental setup of tags and reader antennas that we use to test TagSee.
We deploy a total of $K = 116$ tags on a wooden shelf (each tag is first pasted on a sticky post-it note which is then attached to the shelf).
We deploy $k_x=29$ and $k_y=4$ tags along X and Y axes, respectively, with an inter-tag distance of $5~inches$ along both axes.
The area between the tags is divided into \textit{voxels} or image pixels, such that there are 5 \textit{voxels} between each pair of tags deployed along both axes.
The \textit{voxels} are shown by red dots in Fig. \ref{fig:setup}.
We place two reader antennas $13.78~ft$ away from the shelf, with an inter-antenna distance of $20~inches$ as shown in Fig. \ref{fig:setup}.
Tags are attached such that the distance of the first row of tags is $4.5~ft$ from the ground.
Both antennas are placed parallel to the shelf, such that the distance of their centers is $4.5~ft$ from the ground.
Moreover, as imaging is based on obstruction of LOS paths between tags and the antennas, we pointed both antennas towards the deployed tags.
We mark 6 item categories on the shelf, where each category is covered by 4 separate columns of tags. 
Note that the exact dimensions of the setup are only required by our analytical imaging approach that derives our DNN based approach and serves as its comparison metric.
Our DNN based approach just requires that the locations of tags and antennas do not change after training as that will require retraining the DNN.

\presec\presec
\subsubsection{Data Collection}
For data collection, we recruited 10 users who volunteered to provide data for our project.
From 5 of those users, we were able to collect a total of 14617 samples ($2500+$ samples per person).
We use this data to train TagSee's DNN model because we assume that it is big enough to capture the diversity of browsing movements of those users (who had different body widths and heights) reasonably well. 
However, due to time limitation, the remaining 5 users could only provide us with 2413 samples ($<500$ samples per person).
As the data obtained from these users is limited, we use their data for testing TagSee's performance.
Hence, we test TagSee's performance using unseen data (i.e. data from the users who are not used for training TagSee's DNN), which makes our evaluation more robust. 
%
%
Note that the users in our study had different body widths and heights.
However, an evaluation of the impact of such variations on TagSee's performance is out of the scope of this paper and left as part of future work.
In this paper, we only focus on coarse-grained imaging of customers in front of the shelves. 

\presec\presec\presec
\subsubsection{Performance Metrics}
Except for the scenarios where we compare TagSee's imaging performance, we evaluate TagSee's popularity tracking performance for any experiment using TPRs, FPRs and Miss Rates (MRs), which are calculated based on the correctness of item popularities $\mathcal{P}_j$ TagSee determines in different time windows. 
True positives correspond to the scenarios during which TagSee is able to detect interest in the categories being tested. 
However, TagSee may wrongly detect interest in categories (i.e. other than the ones being tested) as well, which correspond to false positive scenarios. 
TagSee \textit{misses} when it is unable to detect interest in the tested categories during a time window (i.e. MR = 1 - TPR).
Our goal is to achieve maximum TPRs and minimum FPRs.

\presec\presub
\subsection{Single Person Imaging Scenarios}
\postsub
\textit{TagSee can achieve TPRs of more than 90\% and FPRs of less than 5\% for single person monitoring scenarios.}
%
%
Figure \ref{fig:user7testsignleperson} compares the imaging results of these two approaches, where TagSee constructs images of a user as he stands in front of each different item category, using 2 antennas. 
For this experiment, TagSee used a DNN trained for 3 volunteers, where the selected volunteers did not include the tested user. 
We can see that the images constructed by DNN based RFID imaging (bottom) are highly accurate as compared to the ones constructed using baseline approach (top). 
This is because, first, our DNN approach automatically tunes all values in the image construction matrix for minimum construction errors.
Second, the DNN based approach is less vulnerable to natural human motions during browsing activity, which happens because it takes such variations due to motion into account during the training process.
Next, we show how TagSee's performance in single person scenarios is affected by number of training users, impact width and number of antennas.

\presec\presec
\subsubsection{Effect of the number of training users}
Figure \ref{fig:signlepersontraining} shows the impact of number of training users on TPRs, FPRs and MRs for 3 different experiments, performed for 3 of the 5 volunteers selected for testing.
Impact width was set to $k_{cw}=6$, and the error rates reported were averaged over 6 different item categories.
Moreover, data from 2 RFID reader antennas was used for constructing images in these experiments.
We can observe an increasing trend in TPRs for all three users \ref{fig:signlepersontraining1}-\ref{fig:signlepersontraining2}, which is intuitive.
This happens because as TagSee's DNN is trained with more scenarios, corresponding to different users of different shapes and sizes, its image construction becomes more accurate and robust, leading to higher detection rates.
This is also the reason behind the decreasing trend in MRs, which we can observe for all three users \ref{fig:signlepersontraining1}-\ref{fig:signlepersontraining2}.
However, we see that FPRs do not show an expected decreasing trend, which may seem counter-intuitive at first.
This happens because when a customer is browsing an item category, our multi-person imaging algorithm sometimes wrongly detects and images human presence in front of nearby item categories as well, which leads to spurious popularity counts.
%
In this scenario, TagSee can achieve TPRs of more than 85\% and FPRs of less than 15\%, averaged over all users and categories.

\presec\presec
\subsubsection{Effect of the number of reader antennas}
In the aforementioned experiments, we were using 2 reader antennas.
Figure \ref{fig:singleperson1antenna} shows the error rates achieved using single reader antenna, for $k_{cw}=8$ and different number of training users.
We observe that the trend in TPRs and MRs remains similar to the ones corresponding to the 2 antenna scenarios (Fig. \ref{fig:signlepersontraining1}).
We observe that the average FPR drastically increases to more than 45\%.
This is because, 
excluding images from one of the antennas leads to ineffective filtering (as mentioned in $\S$ \ref{sec:multihuman}) of consecutive image frames, which may contain spurious popularity counts for untested categories. 

\presub\presub\presec
\subsection{Multi-Person Imaging Scenarios} \label{multipersonresults}
\postsub
\textit{TagSee can achieve more than 90\% TPRs, and less than 10\% FPRs for 2 person monitoring scenarios. Moreover, for 3 person monitoring scenarios, TagSee can achieve more than 80\% TPRs, and less than 20\% FPRs.}
%
%
Figure \ref{fig:users5_6_7testmultiperson} shows some selected imaging results for these two approaches, which were constructed using 2 reader antennas, where test users performed browsing activities in front of item category sets \{1,3\}, \{1,6\}, \{3,6\}, \{4,6\}, \{1,4,6\} and \{2,4,6\} respectively.
We can observe that TagSee's DNN based RFID imaging approach produces accurate images even for multi-person monitoring scenarios as well.
%
%
We chose 3 of the 5 test users for TagSee's multi-person performance evaluation.
Also, we kept the number of reader antennas $A = 2$ for robust imaging.

\presec
\subsubsection{Effect of the number of training users}
Here, first we discuss performance for 2 person monitoring scenarios i.e. corresponding to the category sets \{1,3\}, \{1,4\}, \{1,5\}, \{1,6\}, \{3,6\} and \{4,6\}.
Figures \ref{fig:multipersontrain1}-\ref{fig:multipersontrain3} show how TagSee's average performance changes with number of training users, for three different values of $k_{cw}$.
We can observe an increasing trend in TPRs (decreasing MRs). We also see a decreasing trend in FPRs, but just like for single person monitoring scenarios, it is not consistent.
For example, in case of $k_{cw} = 8$, FPRs decrease as training users increase from 1 to 4, but we see an increase in FPRs for the case corresponding to 5 training users.
We attribute such unexpected changes in FPRs to spurious popularity counts.
Second, we discuss performance for 3 person monitoring scenarios i.e. corresponding to the category sets \{1,4,6\} and \{2,4,6\}.
Figures \ref{fig:multipersontrain1_a}-\ref{fig:multipersontrain3_a} show TagSee's average performance for different number of training users and values of $k_{cw}$.
Again, we observe an increasing trend in TPRs, however, the overall TPRs achieved are lower as compared to 2 person monitoring scenarios.

\presec\presec
\subsubsection{Effect of impact width $k_{cw}$}
Figures \ref{fig:multipersonimpactwidth1}-\ref{fig:multipersonimpactwidth3} show performance for the multi-person monitoring scenarios corresponding to all tested item category sets.
By closely observing the figures \ref{fig:multipersontrain1}-\ref{fig:multipersontrain3} and \ref{fig:multipersonimpactwidth1}-\ref{fig:multipersonimpactwidth3}, we can see that although, TPRs increase with $k_{cw}$, but FPRs also increase simultaneously.
For both 2 person (1-6 in \ref{fig:multipersonimpactwidth1}-\ref{fig:multipersonimpactwidth3})) and 3 person scenarios (7-8 in \ref{fig:multipersonimpactwidth1}-\ref{fig:multipersonimpactwidth3})), we observe that TPRs increase with $k_{cw}$ in almost each tested scenario, but FPRs also increase, which happens because in multi-person scenarios, the RSS values corresponding to the tags deployed around the item categories in between two nearby customers, are more aggressively affected, which can fool TagSee's multi-person imaging algorithm into wrongly incrementing the popularity counts of those categories.
Although, TPRs achieved for $k_{cw} =$ 4 and 6 are often lower than for $k_{cw} =$ 8, but FPRs corresponding to those cases are considerably lower.
%

		\presec\presec
\section{Discussions and Future Work}\label{sec:discussions}
\postsec
%
%

\smallskip\noindent\textbf{DNN Architecture.}
The number of neurons per layer, number of layers and the types of layers are the primary hyper parameters of our DNN architecture.
The problem of finding the correct hyper parameters for a neural network is a research problem in itself \cite{bergstra2011algorithms, domhan2015speeding, loshchilov2016cma}.
However, unlike standard practice in deep learning where many DNN architectures are randomly tried, the design of our DNN architecture is grounded on an analytical RFID imaging model that we derive in \S \ref{sec:baseline}.
From this model, we derive useful insights based on which we set those hyper parameters of our DNN architecture (\S\ref{sec:neuralnet}).
Our results show that our model driven architecture achieves reasonably good accuracies and generalizes well for unseen data. 
Other hyper parameters like dropout rate, learning rate, and regularization need to be estimated through trial and error, or a grid search.
However, this search can be done before the model is deployed, so its cost does not affect the runtime of TagSee.
Our DNN architecture can be generalized to tag matrices of different sizes and tag density (as discussed in \S\ref{sec:neuralnet}) by changing the parameters $k_x,~k_y,~p_x,~p_y,$ and $K$.
However, its evaluation is out of the scope of this paper.


\smallskip\noindent\textbf{Reading Rate.} 
Figure \ref{fig:readingrateimpact} shows the impact of reading rate (normalized) on the Miss Rates and False Positive Rates in a 3-person imaging scenario.
The MRs and FPRs were obtained for 50 different experiments that we ran for every plotted reading rate.
We can observe that MRs and FPRs increase as reading rate decreases and vice versa. 
This is because imaging depends on variations in RSS signal from all the tags being affected by an obstruction.
Receiving signal from only a few tags leads to imaging inaccuracies resulting in higher MRs and FPRs.
However, note that our goal is to monitor the `browsing' behavior of customers (i.e. when they stop to look at an item category without touching any items), not to continuously track the location of customers as they move about in the store.
Browsing is a slow activity as it assumes that when a customer is browsing some item category, they usually spend a few seconds (e.g. 3-4 seconds) to browse the category.
Because average tag read rate of our current system is approximately 475 reads/second, and there are only 116 tags in our current deployment, our system can obtain enough readings to construct reasonably accurate images of the users standing in front of different item categories.
TagSee can work well in small stores (e.g. a small shoe store) with where the number of shelves is small and the number of deployed tags is on the order of reading rate.
However, for a larger store, we can divide the store into multiple smaller regions and then deploy separate readers to monitor customer activity in each region to ensure that each tag is read frequently. 
In each region, the transmit power of antennas and the frequencies of operation can be set such that the inter-region interference is minimized.

\begin{figure} [htbp]
		\vspace{-0.12in}
	\centering
	\captionsetup{justification=centering}
	\includegraphics[width=0.42\textwidth]{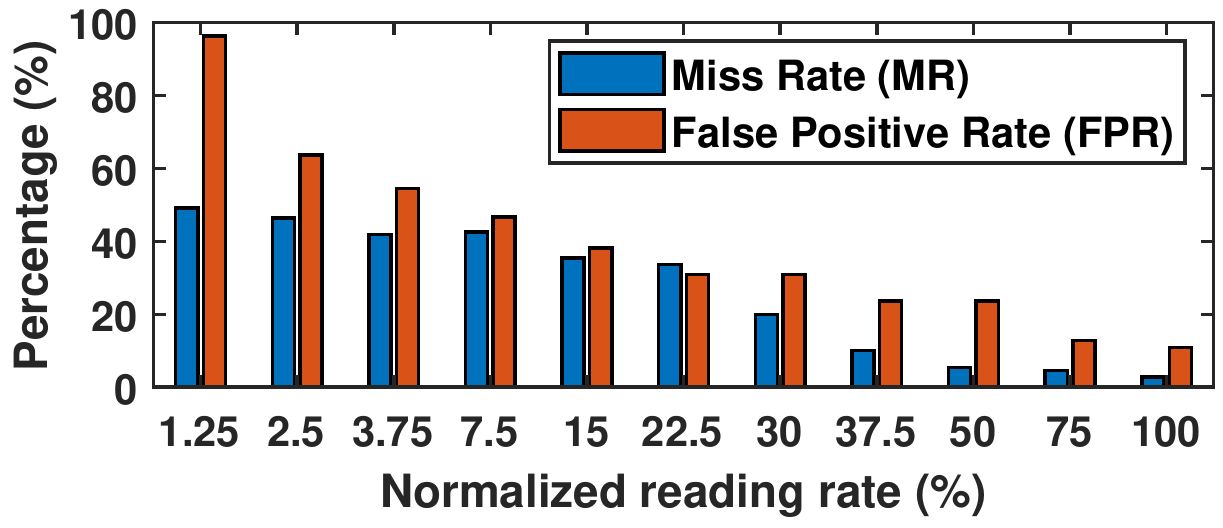}
	\vspace{-0.05in}
	\caption{Impact of reading rate on FPRs and MPRs} 
	\label{fig:readingrateimpact}
	\vspace{-0.12in}
\end{figure}

\smallskip\noindent\textbf{Density of tags and image resolution:} 
The parameters $k_x,~k_y,~p_x,~p_y,$ and $K$ collectively control the size and density of a deployed mesh of tags.
Imaging resolution will naturally be higher if tag density in a mesh is higher (i.e. inter-tag distance is small).
For example, deploying tags more densely along X-axis can help better resolve two customers standing close to each other.
We leave the evaluation of such dynamics between tag density and imaging resolution as part of future work.

\smallskip\noindent\textbf{Collection of training data:} 
Due to time and scheduling constraints, we were only able to collect data from a limited number of volunteers for testing.
Although the data we collected using current setup is enough to test the basic working of TagSee, yet it is just the first step.
In a real-life deployment scenario, for example in a shoe store, an automated camera triggered labeling system can be developed to calibrate TagSee over a period of few weeks. 
Such an automated calibration system will help generate a bigger and more diverse dataset that can be used to train a more robust DNN for TagSee.

\smallskip\noindent\textbf{Practical real-life deployments:}
TagSee's imaging scheme is based on the obstruction of LOS paths, or, more precisely, the Fresnel zones \cite{hristov2000fresnal} between tags and reader antennas.
In our current setup, the tags and antennas are placed such that an approximate LOS is established between them.
However, LOS can also be established in real-life deployments, by hanging the antennas at an angle from the roof using ceiling mounts and attaching the tags to lower racks of the shelves as well as on the floor area near the shelves.
When customers come near a shelf to browse an item category, they would obstruct the Fresnel zones between the tags and their respective reader antennas, based on which TagSee will try to determine the popularity of that item category.

\smallskip\noindent\textbf{Impact of mutual coupling between tags:} The read range of UHF tags can decrease when they are close to each other (e.g. within $100~mm$) due to mutual coupling \cite{tanaka2009change}. 
In our current setup, the inter-tag distance is $127~mm$. However, denser tag deployments can significantly increase coupling and impact read range of the tags.
We leave the evaluation of such mutual coupling dynamics as part of future work.
		\presec\presec
\section{Conclusions} \label{sec:conclusion}
\postsec
In this paper, we propose, implement, and evaluate TagSee, which is the first monostatic RFIDs based imaging scheme, which can be used to monitor browsing activity of customers in places such as physical retail stores. 
%
%
%
To achieve this, we propose a DNNs based imaging approach, which robustly images the browsing activity of customers in front of the shelves with high accuracy.
Our approach is driven by an analytical model, where we first mathematically formulate the problem of imaging humans using monostatic RFID devices and derive an approximate analytical imaging model that correlates the variations caused by human obstructions in the RFID signals.
Afterwards, we use that model to design our DNN's architecture. 
Finally, based on our DNN based approach, we develop a technique which can monitor activity of multiple customers, showing interest in multiple different item categories, simultaneously.
%
%
%
%
The key contribution of this work is in demonstrating the possibility of effective imaging of the browsing activity of multiple customers using existing RFID devices and protocols.
We believe our proposed scheme will help to shorten the gap between online and physical shopping.
	}
	
	\IEEEdisplaynontitleabstractindextext
	\IEEEpeerreviewmaketitle
	
	{ 
		\bibliographystyle{unsrt}
		\bibliography{event}
	}
	
	%
	%
	
\end{document}